\documentclass{article}
\usepackage{times}
\usepackage{amsmath}
\usepackage{mathptmx}
\usepackage{amsfonts,amsbsy}
\usepackage{color} 

\usepackage{graphicx}
\graphicspath{%
    {converted_graphics/}% inserted by PCTeX
    {/}% inserted by PCTeX
}

\begin{document}
%Steve's definitions start here
\def\<{\langle}
\def\>{\rangle}
\def\lap{\nabla^2}
\newcommand{\er}[1]{$(\ref{eq:#1})$}
\newcommand{\sr}[1]{$(\ref{sec:#1})$}
\newcommand{\ssr}[1]{$(\ref{ssec:#1})$}
\newcommand{\fr}[1]{$(\ref{fig:#1})$}
\newcommand{\bg}{\begin{equation}}
\newcommand{\ee}{\end{equation}}
\newcommand{\lla}[1]{\label{lem:#1}}
\newcommand{\tla}[1]{\label{thm:#1}}
\newcommand{\ela}[1]{\label{eq:#1}}
\newcommand{\xiv}{\bm{\xi}}
\newcommand{\ov}{\overline}
\newcommand{\mach}{\sqrt{M^2-1}}
\newcommand{\app}{A^{\prime\prime}}
\newcommand{\signof}{\rm sgn}
\newcommand{\siga}{{\sigma_\alpha}}
\newcommand{\iv}{{\bf i}}
\newcommand{\uv}{{\bf u}}
\newcommand{\nv}{{\bf n}}
\newcommand{\xv}{{\bf x}}
\newcommand{\av}{{\bf a}}
\newcommand{\ev}{{\bf e}}
\newcommand{\Fv}{{\bf F}}
\newcommand{\Bv}{{\bf B}}
\newcommand{\Uv}{{\bf U}}
\newcommand{\rey}{\mathit{Re}}
\newcommand{\Jv}{{\bf J}}
\newcommand{\Av}{{\bf A}}
\newcommand{\gv}{{\bf g}}
\newcommand{\hv}{{\bf h}}
\newcommand{\vv}{{\bf v}}
\newcommand{\kv}{{\bf k}}
\newcommand{\tv}{{\bf t}}
\newcommand{\bv}{{\bf b}}
\newcommand{\Rv}{{\bf R}}
\newcommand{\qv}{{\bf q}}
\newcommand{\rv}{{\bf r}}
\newcommand{\yb}{\bar{y}}
\newcommand{\jv}{{\bf j}}
\newcommand{\Qv}{{\bf Q}}
\newcommand{\Vv}{{\bf V}}
\newcommand{\eps}{\epsilon}
\newcommand{\para}{\parallel}
\newcommand{\del}{\nabla}
\newcommand{\ptl}{\partial}
\newcommand{\om}{\omega}
\newcommand{\yv}{{\bf y}}
%Steve's definitions end here

%Andrew's definitions begin
\newcommand{\fdot}{\dot{f}}
\newcommand{\gdot}{\dot{g}}

\newcommand{\Green}{\mathrm{Green}}
\newcommand{\harm}{\mathrm{harm}}
\newcommand{\squeeze}{\mathrm{sq}}
\newcommand{\trans}{\mathrm{trans}}
\newcommand{\scale}{\mathrm{scale}}
\newcommand{\contraction}{\mathrm{frame}}

\newcommand{\Gc}{\mathcal{G}}
\newcommand{\Jc}{\mathcal{J}}
%Andrew's definitions end

\newcommand{\nvh}{\hat{\bf n}}
\newcommand{\rvh}{\hat{\bf r}}
\newcommand{\tvh}{\hat{\bf t}}
\newcommand{\zvh}{\hat{\bf z}}
\newcommand{\thetavh}{\hat{\bf \theta}}
\newcommand{\rhovh}{\hat{\bf \rho}}
\newtheorem{lemma}{Lemma}
\newtheorem{corollary}{Corollary}
\newtheorem{theorem}{Theorem}

%\shorttitle{Vorticity growth} %for header on odd pages
%\shortauthor{S. Childress et al} %for header on even pages

\title{Eroding dipoles and vorticity growth \\ for Euler flows in $ \scriptstyle{\mathbb{R}}^3$ \\ I. Axisymmetric flow  without swirl}
\author{Stephen Childress\\New York University\\Courant Institute of Mathematical Sciences\\Andrew D. Gilbert\\University of Exeter\\Department of Mathematics\\and\\Paul Valiant\\
Department of Computer Science\\ Brown Universit}

\maketitle

\begin{abstract}
A review of analyses based upon anti-parallel vortex structures suggests that structurally stable vortex structures with eroding circulation 
may offer a path to the study of rapid vorticity growth  in solutions of Euler's equations in $ \scriptstyle{\mathbb{R}}^3$. We examine here the possible formation of such a structure in axisymmetric flow without swirl, leading to maximal growth of vorticity as $t^{4/3}$. Our study suggests that the optimizing flow giving the $t^{4/3}$ growth mimics an exact solution of Euler's equations representing an eroding toroidal vortex dipole which locally conserves kinetic energy. The dipole cross-section is a perturbation of the classical Sadovskii dipole having piecewise constant vorticity, which breaks the symmetry of closed streamlines. The structure of this perturbed Sadovskii dipole is analyzed  asymptotically at large times, and its predicted properties are verified numerically. 

\end{abstract}

\section{Introduction}

The purpose of this paper is to outline an approach to the construction of an Euler flow involving an eroding dipole structure which achieves maximal growth of vorticity in axisymmetric flow without swirl.  Our analysis will employ asymptotic estimates and neglect certain higher-order effects, but the results will be supported by numerical calculations. Our aim is to present a plausible if approximate physical model with a number of compelling features,  which enables some explicit (if formal) analyses of vorticity growth in three dimensions.

We focus here on the \textit{local} amplification of vorticity, in other words on the \textit{self-stretching} of a vortex structure. This is in contrast to the stretching that results from
distant interactions of vortex structures. The  scaling invariance inherent in Euler flows allows such local stretching to proceed in principle to arbitrarily small scales, allowing extremely rapid growth.  This viewpoint has indeed motivated much of the research into the possibility of blow-up of vorticity in finite time in three dimensions, and has led almost exclusively to consideration of the interaction of \textit{anti-parallel} vortex structures. An excellent summary of this research may be found in  \cite{Gib}. We mention in particular the work of \cite{PS} on the interaction of anti-parallel, thin vortex tubes, research which introduced the possibility of a finite-time singularity. However it turns out this interaction cannot avoid the ultimate distortion of vortex cores. This is because, considered as line vortices, the motion brings the  filaments together at a rate which is proportional to the  logarithm of the product of the radius  and the curvature. This product must remain large to ensure the integrity of the cores. This then implies that the distance between the filaments shrinks faster than the core size, so distortion must occur and the filament model fails; for details see \cite{HB}.

Explicit numerical studies of core interaction in three dimensions have again involved anti-parallel tubes, see \cite{Gib}. We mention in particular the work of  \cite{BK} and  \cite{HL}. The problem of core interaction also occurs in the simpler problem of collision of two vortex rings, see \cite{osh}, \cite{LN}, and \cite{Ri}. This brings us to one main focus of this paper, the interaction of anti-parallel vortex tubes in axisymmetric flow without swirl (AFWOS).
In a subsequent Part II, we shall extend the discussion to include general anti-parallel structures in three dimensions.

Our approach in the present paper is to use this simpler problem to explore in detail core interaction. It is well known that in AFWOS there can be no finite time blow-up of vorticity ( \cite{MB}. Nevertheless this is an arena where modest amplification of vorticity can be studied in detail. We have argued that this problem leads naturally to the important role played by the  local conservation of total kinetic energy ( \cite{Ch1}). This enforces a loss of volume of the vortical structure associated with growth, which can then be described as an ``eroding'', toroidal,  dipolar structure. 
Such erosion is sometimes also described as ``stripping'', in which a vortex loses outer layers of vorticity, thus sharpening the vortex profile. 
We show that such a structure should emerge generally from  equal and opposite colliding vortex rings, and the ultimate fate can be realized by a solution of Euler's equations corresponding to a eroding, locally two-dimensional structure having a uniform vorticity in each of the two constituent eddies. The non-eroding counterpart is the well-known 2D Sadovskii vortex with continuous velocity ( \cite{Sad, PH, ST}). 

There is already clear numerical evidence for the existence of such solutions. Studies of interactions of anti-parallel vortex rings have suggested that vorticity tends to be shed into a sort of ``tail'' aft of the main body of the resulting dipolar vortex, as the tubes are stretched, see e.g. \cite{SMO}, \cite{BK}, and \cite{GR}. Calculations of colliding rings using the techniques of contour dynamics explicitly exhibit the development of a long ``tail'' and a ``tadpole'' shape for the dipole/tail structure; see \cite{Ri}, \cite{SLF}. In  \cite{SLF} it was shown that the head of the tadpole is indeed very close to the shape of the  Sadovskii dipole. Our claim here is that this configuration emerges generally in AFWOS under the condition that we are dealing with a toroidal dipole that is anti-symmetric about a plane dividing the two vorticity regions.  Note that the physical experiment described in  \cite{LN} involves vortex stretching in an apparently axisymmetric phase, before an non-axisymmetric instability develops that ends the expansion.\footnote{A video may be seen at http://www.youtube.com/watch?v=12ozAloKYyo.} 
However prior to the instability one observes  an axisymmetric ``membrane'' which is consistent with the shedding of a tail behind the axisymmetric dipole pair propagating radially outwards. 

The paper is organized as follows. Section 2 introduces the axisymmetric geometry, and the origins of the basic scalings of the eroding dipolar structure . In Section 3 we present a simplified analysis of the eroding dipole based upon the \emph{ad hoc} temporal scalings of size and velocity derived in section 2 from the constraint of locally constant kinetic energy. In Section 4 the program outlined in Section 3 is subjected to more detailed asymptotic analysis in order to  compute these scalings explicitly. We shall thereby derive the property of local energy conservation directly from the dynamics of a perturbed Sadovskii vortex. In section 5 numerical simulations are described which are found to exhibit the scalings of the previous sections as well as show the evolution toward the asymptotic state. 

\section{Preliminaries}

\subsection{The axisymmetric geometry} 

In  AFWOS, we may study the growth of vorticity  by expansive stretching  in its simplest setting. It is known that no finite time singularity can then be formed \cite{MB}), but one may still pose an initial value problem in $\mathbb{R}^3$ and ask how fast vorticity can grow at large times. The problem was taken up in \cite{Ch1}, and we now summarize the results. The ideas outlined here will be developed  further in  section 2.

Since the vorticity consists of rings with a common axis, maximal growth can be determined by considering a symmetric anti-parallel bundle of vortex rings. Optimization under the condition of conservation of vorticity volume then bounds the maximum of vorticity as a multiple of $t^2$. This estimate can be understood as follows. Imagine a torus with a  centreline $\mathcal{C}$ of radius $R$,  and a { circular} cross-section of  radius $a$. Let the angular vorticity component (the only component to be considered here) be $\pm \omega_\theta$ in the two half-discs of the cross-section, the signs such as to produce expansive stretching. By conservation of volume, $a^2R \sim 1$ in order of magnitude. Also, vortex dynamics ensures $dR/dt\sim \omega_\theta a$. Finally, conservation of vorticity flux requires $\omega_\theta a^2\sim 1$. Thus $dR/dt\sim \sqrt{R}$ leading to the $t^2$ estimate. We now set $\omega_\theta =\omega$ for the axisymmetric case.

This bound is not sharp since the optimizer does not conserve total kinetic energy $E$. Indeed $E\sim Ra^2 (\omega a)^2\sim R$. If conservation of energy is also imposed, and a cross-sectional scale determined again by a single length $a$, one must take $a\sim R^{-3/4}$ so that vorticity volume is lost. In fact we can only maintain kinetic energy approximately, with loss of volume and energy occurring through the shedding of a ``tail'' of vorticity laden fluid from the vortex pair, of thickness $H\sim R^{-5/2}$.

To see this, suppose then that we seek a structure with $a\sim R^{-p},\; p > 0$, which extrudes a tail in the form of a sheet of thickness $H\sim R^{-q}$. Conservation of volume requires that
\bg
a\dot{a} R + a^2\dot{R} \sim R\dot{R} R^{-q}
\quad \implies\quad
 R^{-2p} \dot{R} \sim  R^{1-q} \dot{R},
 \label{eqEhead} 
\ee
so that $q = 2p+1$. For the kinetic energy, considered relative to the fluid at infinity, energy in the tail is created at a rate
\bg
\frac{d E_{\mathrm{tail}}}{ dt} \sim R \dot{R} (\omega R^{-q})^2 R^{-q} \sim R^{ 3-3q} \dot{R}.
\label{eqEtail} 
\ee
If this must equal the energy decrease in the ``head'' of the structure, estimated as
\bg
\frac{d E_{\mathrm{head}} }{dt}  \sim {d\over dt}\, [Ra^2(\omega a)^2]\sim{d\over dt} \, R^{3-4p}\sim (3-4p) \dot{R} R^{2-4p},
\ee
then $R^{2p + 1} \sim 1$, which is impossible. The only recourse is to set $3-4p = 0$,  to make $E_{\mathrm{head} } \sim 1$. Thus $(p,q)=(3/4,5/2)$ and $dR/dt\sim \omega R^{-3/4}\sim R^{1/4}$, yielding a maximum growth as $R\sim t^{4/3}$.  Kinetic energy is lost to the tail at a rate ${ R^{1-7p}=R^{-17/4}}$ from (\ref{eqEtail}) and so is extremely small at large $R$, consistent with $E_{\mathrm{head}} \sim 1$ in (\ref{eqEhead}).

\begin{figure}
  \centering
  \includegraphics[bb=0 0 439 356,width=2in,height=1.62in,keepaspectratio]{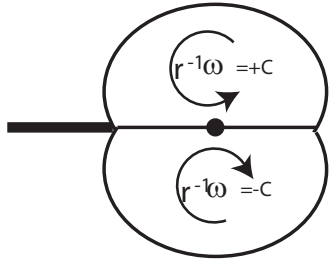}
  \caption{Vorticity distribution for the dipole which maximizes the velocity of the centreline ``target'' (the centre dot), subject to constraints on volume, $\omega/r$, and total kinetic energy. Here $C$ denotes the maximum initial value of $|\omega/r|$. The trailing black line represents the thin sheet of extruded vorticity, conserving volume with negligible loss of kinetic energy.}

  \label{fig:optimizer}
\end{figure}

In \cite{Ch1} we described the solution to the variational problem for conserved energy and volume. Details, and related work without energy conservation, are given in \cite{Ch3}. The form of the optimizing dipole is shown in figure~\ref{fig:optimizer}. The extruded ``tail'' conserves volume while negligibly reducing kinetic energy.

\begin{figure} 
\centering
\vskip -2in
\hskip -3in
  \includegraphics[bb=0 0 1229 820,width=5.67in,height=3.78in,keepaspectratio]{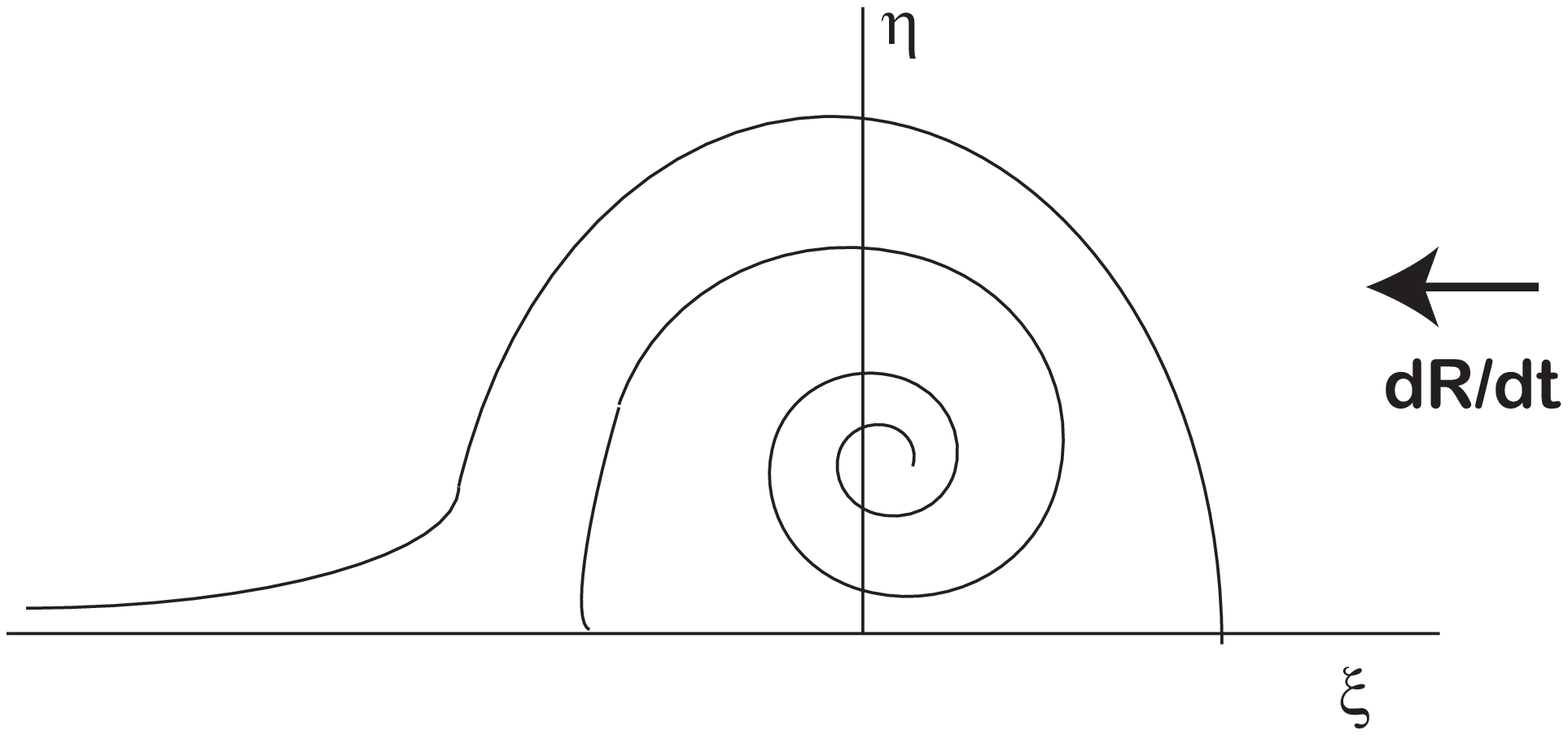}
  \caption{The upper half of the snail, showing flow lines relative to a comoving frame, in local coordinates $(\xi,\eta) = R^{3/4}(x,y)$. The spiral shown is the unique flow line terminating at a stagnation point on $\eta=0$.}
  \label{fig:asympsingfig2}
\end{figure}

Despite their origin from a problem with axial symmetry, these last estimates provide a crucial piece of information concerning the structure of fast-growing vortical structures of this kind.  For an anti-parallel, symmetric pair of adjacent eddies, conservation of energy forces a contraction of eddy cross section over  and above that imposed by conservation of volume. Relative to a co-moving frame with coordinates suitably normalized (here by a factor $R^{3/4}$), the apparent flow now contains a small non-solenoidal component, effectively feeding volume into the tail as the true cross-section contracts. The key point is that this component breaks the constraint of closed streamlines that prevails without it. In effect conservation of energy turns a structurally unstable topology into a
structurally stable, spiral topology.  We sketch the proposed flow lines of the upper eddy, a structure we shall refer to as the ``snail'', in Figure~\ref{fig:asympsingfig2}. (Note that the direction of motion of the vortices is opposite to the direction of crawling of the ``snail''.) 

\subsection{Remark on the constancy of kinetic energy}

Although we have argued that the approximate constancy  of kinetic energy implies a certain asymptotic dipole structure, we have not excluded the possibility that dipole kinetic energy is in fact
{\em not} conserved. That is, can a finite kinetic energy be lost to the tail along with volume? We now consider this possibility.

Again let $a\sim R^{-p}$, and  kinetic energy of order $R^{3-4p}$. The question is, can $p$ exceed 3/4? We saw in section 2.1  that the flux of kinetic energy into the tail is of order  
$R^{1-7p}$. 
Let us determine kinetic energy as a function $F(R)$ which would yield this flux. Then, since $dR/dt\sim R^{1-p}$,
\bg
F^\prime (R) R^{1-p}\sim R^{1-7p},
\ee
so that $F = F_0+ f(R)$ where $F_0$ is a constant and $f\sim R^{1 -6p}$. But if $p>3/4$ we know that kinetic energy is decreasing as $R^{3-4p}$. Thus $F_0=0$ and $1-6p= 3-4p$, which is impossible if $p > 3/4$. Note that if $p=3/4$ a similar argument allows $F_0 > 0$ and $f\sim R^{-7/2}$.
Thus we are forced to essentially maintain kinetic energy and $p=3/4$ stands as the exponent determining the only viable dipole structure.
\subsection{The change of topology in AFWOS}

We now summarize results to be derived in the next section. An asymptotic analysis of the maximal growth of vorticity in time results in the leading order flow and vorticity field $\omega$ taking the form of an arbitrary two-dimensional eddy with closed streamlines. 
At the next order, however, contour averaging introduces a compatibility constraint on the leading vorticity term, needed to ensure the existence of the solution at second order.

We are dealing, therefore, with a singular perturbation in the topology of the flow. Realizing that the actual structure at second order is that of the snail, we see that the compatibility constraints on individual closed streamlines are removed,  provided that the perturbed velocity field is taken as the ``leading flow field'',  establishing the spiral flow lines.  Thus we have an example of bringing forward a second-order effect, in order to completely reorder a calculation, here for the purpose of correctly identifying the spiral topology. Once this is done, flow along spiral flow lines terminating in the tail determines the tail, that is, determines  the vorticity shed to the wake, irrespective of the particular form of the forcing at second order.

Analysis of the snail configuration will lead naturally to the hypothesis that the preferred ultimate vorticity distribution is one which is piecewise constant, corresponding to zeroth-order (unperturbed) eddies which are of the form of the Sadovskii vortex. We use ``the'' here to refer to the limiting dipole configuration where velocity is continuous at the boundary. This dipole is one of a family, allowing a discontinuity of velocity at the boundary, considered by  \cite{Sad}. The solution of interest here was independently studied by  \cite{ST}  ; for a review of these problems see \cite{MST}. 
The constant vorticity regions emerge from any other dipolar configuration by the stripping away of vorticity into the tail, leaving tubular neighbourhoods of two symmetric anti-parallel vortex lines, thus giving the Sadovskii structure. 

\section{Analysis of vorticity growth in AFWOS}

We now turn to the asymptotic analysis of a dipolar vortex structure under the constraints of AFWOS, for large values of $R$. Motivated by the preceding estimates and bounds, we shall seek a structure whose cross-sectional area decreases as $R^{-3/2}$ but which maintains self-similarity of shape to leading order. To contrast the present discussion with the analysis to be presented in section 3, our point now is to impose explicitly the scaling associated with $R\sim t^{4/3}$ and a dipole area decreasing as $t^{-2}$ and to assess the resulting equations for large $t$. In fact we shall consider a slightly more general class where area goes as $R^{-2p}$, and identify the scaling which preserves kinetic energy with the value $p=3/4$.

\subsection{Local analysis of eroding dipoles}

In cylindrical coordinates $(r,z,\theta)$ (this order being chosen as we shall be working largely in the $(r,z)$-plane),  the vorticity equation is
\bg
\left[ {\ptl\over\ptl t} + u_r\, {\ptl\over\ptl r} + u_z\, {\ptl\over\ptl z }\right] {\omega \over r} = 0,\quad \omega ={\ptl u_z\over\ptl r}-{\ptl u_r\over\ptl z}\, .
\ela{exact1}
\ee
The conservation of volume is expressed, for an incompressible fluid of unit density, by
\bg
{1\over r}\, {\ptl r u_r\over\ptl r}+{\ptl u_z\over \ptl z} = 0.
\ela{exact2}
\ee
We now pass to local coordinates $(x,y)$ though the transformation $r= R(t) + x$, $z= y$, $u_r=\dot{R}+u$, $u_z=v$, with $\dot{R} = dR/dt$. We then have
\bg
\left[ {\ptl\over\ptl t} + u\, {\ptl\over\ptl x} + v \,{\ptl\over\ptl y }\right] {\omega\over R(t) +x} = 0,\quad \omega = v_x-u_y,
\ee
\bg
u_x+v_y +{u\over R(t)+x}= -{\dot{R}\over R(t)+x}\, .
\ee
The time derivative is now understood to be for $x$ fixed.

Writing
\bg
\omega = \omega_0 \, {R(t)\over R_0} \, \Omega(x,y,t),
\ee
where $R_0$ is a reference length and $\omega_0$ a reference vorticity, for example associated with initial conditions,  we have
\bg
\left[ {\ptl\over\ptl t} + u \, {\ptl\over\ptl x} + v \, {\ptl\over\ptl y }\right] {\Omega R\over R+x}  = 0,
\quad 
u_x+v_y +{u\over R+x}= -{\dot{R}\over R+x}\, . 
%\quad 
%\dot{R} \equiv  \frac{dR}{dt} \, .
\ee
Next, let $a(t)=R_0 (R(t)/R_0)^{-p}$ be a lateral scale for the dipole, and set $(\xi,\eta)=a^{-1}(x,y)$. Here $p$ is an exponent we expect to be $3/4$ from the previous section, but we leave unspecified for the moment as we wish to emphasize the independence of the constraint of conservation of energy from the form of the dipole topology. We may assume for eroding vortices that $p > 1/2$ (so volume $a^2R$ decreases). Lastly, we set
\bg
(u,v) =\omega_0 R_0 \left({R\over R_0}\right)^{1-p}(U,V)- p \, \left({\dot{R}\over R }\right)  (x,y),
\ee
and define a dimensionless time $\tau$ by
\bg
{\ptl \tau\over\ptl t} = \omega_0 \, \frac{R}{R_0}\, .
\ee

In these variables we set $h=1+x/R$ and have
\bg
\left[{\ptl\over\ptl \tau}+U\,{\ptl\over\ptl \xi}+V\, {\ptl\over\ptl \eta}\right]{\Omega\over h} = 0,\quad U_\xi+V_\eta + {\eps U\over h}= {(2p-1)\eps\over h}\, .
\ela{nondim}
\ee
Here we have chosen $\omega_0,R_0$ so that
\bg
\dot{R} =\omega_0 R_0 (R/R_0)^{1-p},\quad R=R_0 (1+p\omega_0 t)^{1/p},
\ee
and therefore
\bg
\eps=\left({R_0\over R}\right)^{1+p}= {\dot{R} R_0\over \omega_0 R^2}=  {R_\tau\over R}=\eps(\tau)= {1\over 1+(1+p) {\tau}}\sim (1+p)^{-1}\tau^{-1},\quad h = 1+\eps\xi . 
\ee

One useful point to note is that  the tail thickness here is $R^{-(1+2p)}$ and so  the vorticity in the tail contributes a velocity of order $R\times R^{-(1+2p)}\sim R^{-2p}$, whereas relative to the head the fluid flow exits the tail with velocity of order $R^{1-p}$ so as to match with the free stream velocity, consistent with the above estimates. Thus the vorticity in the tail contributes a velocity which is negligible compared  to the free stream.

\subsection{Formal expansion}

We now return to \er{nondim} and carry out a formal expansion in $\eps$ (or  $\tau^{-1}$) for the flow in the upper half of the dipole. We introduce the new time variable $\tau^*$ by
\bg
{\ptl \Omega_0 \over \ptl \tau} = \eps \, {\ptl\Omega_0\over \ptl \tau^*}\, . 
\ee
Thus $\tau^*= (1+p)^{-1} \log[1+(1+p)\tau]\sim (1+p)^{-1} \log \tau$.
Then \er{nondim} becomes
\bg
\left[\eps\, {\ptl\over\ptl \tau^*}+U\, {\ptl\over\ptl \xi}+V\, {\ptl\over\ptl \eta}\right]{\Omega\over h} = 0,\quad U_\xi+V_\eta + {\eps U\over h}= {(2p-1)\eps\over h}.
\ela{nondim2}
\ee
The reasoning here is that the scaling of the coordinates and velocity components has already absorbed the dominant time dependence, and we are left with the slower dependence on $\tau^*$.
Of course now $\eps$ may be regarded as a function of $\tau^*$, with
\bg
{\ptl\eps\over\ptl \tau^*}=-(p+1)\eps.
\ee

We now  let  $\Qv \equiv (U,V)= \Qv_0(\xi,\eta,\tau^*)+\eps\Qv_1(\xi,\eta,\tau^*)+\dots$ and $\Omega=\Omega_0(\xi,\eta,\tau^*)+\eps\Omega_1(\xi,\eta,\tau^*)+ \dots$. We then obtain the equations
\bg
\Qv_0\cdot\nabla \Omega_0 = 0,\quad \nabla\cdot\Qv_0 = 0,
\ela{eq0}
\ee
\bg
{\ptl \Omega_0\over \ptl\tau^*}+\Qv_0\cdot\nabla \Omega_1+\Qv_1\cdot\nabla\Omega_0 -U_0\Omega_o = 0,\quad \nabla\cdot\Qv_1=2p-1 - U_0.
\ela{eq1}
\ee
Let us first solve \er{eq0} simply by setting $\Omega_0 =$ constant  in a lobe of the vortex. Then at next order a particular solution is seen to satisfy
\bg
\Omega_1 = {\ptl V_1\over\ptl \xi}-{\ptl U_1\over\ptl\eta} = \Omega_0 \xi,\quad
 {\ptl U_1\over\ptl \xi} + {\ptl V_1\over\ptl \eta} = 2p-1 - U_0.
\ela{geneq1}
\ee
For example, we can take 
\bg
U_1= \tfrac{1}{2} \eta V_0+ (p- \tfrac{1}{2}) \xi,\quad 
V_1=(p- \tfrac{1}{2}) \eta+ \tfrac{1}{2}\Omega_0 \xi^2) + \tfrac{1}{2} ( \Psi_0-\eta U_0)- \tfrac{1}{4} \eta^2\Omega_0, 
\ela{geneq2}
\ee
where the streamfunction $\Psi_0$ is specified by 
\bg
(U_0,V_0)= \left(-{\ptl\Psi_0\over\ptl\eta},{\ptl\Psi_0\over\ptl \xi}\right). 
\ee
Any potential flow can be added to this solution and the result matched with an exterior potential flow to make the velocity continuous on the bounding streamline of the vortex.
%%
%
%\emph{\color{red} is it helpful here to refer to a sketch or otherwise of the S vortex? Figures 1 and 2 are not quite what is wanted. Maybe we should bring forwards figure 5(a) with a schematic sketch also? If you say what you think, I can sort it out if you wish. I can imagine a reader flailing a bit without some feeling for what the S vortex actually looks like, to fix ideas. I know this is there in the next sentence...}  
%
The point is then that we have a way of extending the zeroth-order solution. In fact we know that there exists a $\Qv_0$ of the desired form, namely the Sadovskii vortex with continuous total pressure.

\subsection{The general case}

We shall say that the dipole vortex is {\em compatible} if an asymptotic solution exists inclusive of the terms of order $\eps$.
We now establish a simple but somewhat surprising result:

\begin{lemma}
The class of eroding dipoles just studied, where vorticity is constant in each eddy, is the only compatible, zeroth-order flow field independent of $\tau^*$.
\end{lemma}

\noindent
To prove this we note from \er{eq0} that the general solution has the form $\Omega_0 = F(\Psi_0)$ where $F$ is an arbitrary function,
Then, from \er{eq1} we see that
\bg
\Qv_0\cdot\nabla \Omega_1+\Qv_1\cdot\nabla \Psi_0 \, F^\prime(\Psi_0)= U_0 \, F(\Psi_0).
\ela{gen}
\ee
Note that strictly the function $F(\psi_0)$ here has several branches for the Sadowskii vortex, because of the two regions of closed streamlines and the one region of open streamlines. So we are working within the region of space for one of these branches. 
We introduce the contour average
\bg
\< \cdot  \> = \oint {\cdot\over |\Qv_0|}\, ds,
\ee
taken along the direction of flow around a streamline of the flow $\Qv_0$ in the upper eddy. Then it is easy to see from \er{gen} that
\bg
\<\Qv_0\cdot\nabla\Omega_1\> = 0 = F^\prime(\Psi_0)\oint\Qv_1\cdot\nv \,  ds+F(\Psi_0)\, \<U_0\>.
\ela{geneq}
\ee
However, from the divergence theorem and \er{eq1},
\bg
\oint\Qv_1\cdot\nv\,  ds= \int\!\!\!\! \int\left( 2p-1 - U_0\right) \, d\xi \, d\eta= (2p-1) \, A(\Psi_0)
\ela{geneq2}
\ee
($U_0$ makes no contribution), where $A$ is area within a contour of constant $\Psi_0$ in the $(\xi,\eta)$ plane. Also 
\bg
\left\<U_0\right\>=\oint{U_0\over |\Qv_0|} \, ds = \oint {dx\over ds} \,  ds = 0.
\ee
It then follows from \er{geneq} and $p>1/2$ that $F^\prime(\Psi_0) = 0$ and the lemma is proved.\footnote{This lemma brings to mind the classical Prandtl--Batchelor result
concerning the constancy of vorticity in steady flow in a region of  closed streamlines at large Reynolds number; see \cite{Batch}. Indeed that work inspired investigation of the associated Euler flows with eddies of opposite sign, representing the wake behind a bluff body translating at a modest Reynolds number ( \cite{Ch5}). However the proof of the Prandtl--Batchelor  theorem uses the Navier--Stokes equation, since the result depends upon the small but persistent diffusion of vorticity , whereas here the drift to the constant state follows from erosion of vorticity in an inviscid flow.}

Now we allow $\Omega_0$ to depend upon $\tau^*$. The contour average then gives
\bg
 \left\<{\ptl\Omega_0\over \ptl \tau^*}\right\>+{\ptl \Omega_0\over\ptl \Psi_0}\oint\Qv_1\cdot\nv \, ds+F(\Psi_0) \, \<U_0\>=\left\<{\ptl\Omega_0\over \ptl \tau^*}\right\>+{\ptl \Omega_0\over\ptl \Psi_0}\, (2p-1)\,  A(\Psi_0).
\ee
But
\bg
\left\<{\ptl\Omega_0\over \ptl \tau^*}\right\>= {\ptl\Omega_0\over \ptl \tau^*}\Big|_{\Psi_0} \< 1\> +{\ptl\Omega_0\over \ptl \Psi_0}\left\<{\ptl\Psi_0\over\ptl \tau^*}\right\>.
\ee
Since (see e.g. \cite{Ch4})
\bg 
\< 1\> = \frac{\ptl A}{{\ptl \Psi_0}} \, , 
\quad
- \left\<{\ptl\Psi_0\over \ptl \tau^*}\right\>= {\ptl A\over\ptl \tau^*}\, ,
\ee
we have
\bg
{\ptl\Omega_0\over \ptl \tau^*} \, {\ptl A\over\ptl\Psi_0}-{\ptl\Omega_0\over \ptl \Psi_0}\left[{\ptl A\over\ptl \tau^*}-(2p-1)A\right] = 0.
\ee
Consequently
\bg
 \Omega_0 = G\left(e^{-(2p-1)\tau^*}A(\Psi_0,\tau^*)\right),
\ee
for some function $G$, But $e^{-(2p-1)\tau^*}\sim \tau^{-{2p-1\over 1+p}}\sim R^{1-2p}$ and  $R^{1-2p}A$ is equal to $R$ times the dimensional area. Thus we obtain a steady flow preserving constant total volume, with area shrinking as $R^{-1}$. This is a compatible dipole for arbitrary $F(\Psi_0)$ since it corresponds to \er{geneq}, \er{geneq2} with $p=1/2$. The  dependence on $\tau^*$ when $p > 1/2$  results from observing a steady volume-preserving structure within  coordinates shrinking faster than is required by conservation of volume.

Using the term ``steady'' in the above sense, meaning that $\Omega_0$ is independent of $\tau^*$,  we thus have the following result:

\begin{theorem}
The compatible dipole vortex structures consist of the steady eroding vortices ($p > 1/2$) with piecewise constant vorticity, and the steady volume preserving vortices ($p=1/2$) with 
arbitrary $F(\Psi_0)$. 
\end{theorem}

We emphasize that compatibility is a fairly weak measure of dynamic consistency, leaving the requirement of constant kinetic energy as an added and independent constraint. The exponent $p$ needs to be fixed by a full asymptotic solution for large $R$ involving matching an eroding vortex to an external potential flow, as well as proper treatment of the vorticity ``tail'', and this requires a numerical solution for the perturbed Sadovskii vortex, a problem we take up in the next section. We know of course that the unique compatible structure preserving total kinetic energy is the steady eroding vortex with $p=3/4$.

In spite of the limited implications of compatibility, we do gain a basic constraint of the zeroth-order structure. We know that the vorticity squared  of the dipole, times the area of one vortex, divided by the speed of propagation squared, must equal 37.11 ( \cite{ST}).  Let the dipole be at position $R\gg R_0$, moving with speed $\dot{R}= \omega_0 R_0(R/R_0)^{1/4}$, and having vorticity $\omega_0 R/R_0$ and area  $2A_0(R_0/R)^{3/2}$. It then follows that
\bg
R_0= \sqrt{A_o/37.11}
\ee 
is our reference length.

We remark that the structure of our preferred dipole with $p=3/4$ can be studied directly in the  stable topology. The idea is simply to take the ``zeroth-order''
velocity  of the snail velocity field, $(U_s,V_s)$ say, to include the apparent fluid source to order $\eps$:
\bg
(U_s,V_s) = (U_0,V_0)+ \tfrac{1}{4} \eps (\xi,\eta),\quad
{\ptl U_0\over\ptl \xi}+{\ptl V_0\over \ptl \eta} = 0,
\ee
where again $(U_0,V_0)$ is the unperturbed Sadovskii velocity field. Our ``zeroth-order'' problem then becomes;
\bg
U_s \, {\ptl \Omega_s\over\ptl\xi} + V_s\, {\ptl \Omega_s\over\ptl \eta} = 0, \;\;
\Omega_s={\ptl V_s\over\ptl\xi}-{\ptl U_s\over\ptl \eta}.
\ee

Our result is now immediate. All flow lines of each vortex are spirals out of a common centre. Since $\Omega_s$ is constant on these flow lines, $\Omega_s$ must be equal everywhere to the value at this centre. We thus may understand the perturbed Sadovskii structure as a result of eroding away the outer layers of the initial structure. 

%\emph{\color{red} we should probably discuss this erosion process qualitatively earlier on in this paper, where we stress the importance of the piecewise constant vortices. Indeed one can even estimate just how constant the vorticity is after a time, given what has been eroded away in the shrinking cross section. Maybe we even need a schematic of the process early on to keep referring to.} 

\subsection{Summary of the composite solution at leading order}

We have seen that the snail emerges as the only compatible vortex structure conserving kinetic energy. We now shall describe the ``leading order'' structure in its entirety, including the external potential flow. By ``leading order'' we here mean that the nominally higher-order effect needed to capture the vortex shrinkage be included as a leading order effect. We thus will describe a perturbed Sadovskii vortex. We are here neglecting entirely the dynamics of erosion. We assume a contracting Sadovskii dipole, at a rate determined by the assumption of energy conservation, and match this with an external flow. To justify this leading-order solution one must {\em derive} the erosion from the equations of motion, and this problem we will take up in the following section.

\subsubsection{The exterior flow}

We begin with calculation  of a uniform approximation to the external potential flow. This flow exists outside the structure consisting of the Sadovskii vortex plus tail. In fact the tail  will not be considered in detail as it will have no active role in the leading order solution.

It is helpful to first consider a simpler potential flow problem, that of an expanding, volume preserving torus centred at $r=R(t)$ with local radius $a(t)$. We present this calculation in appendix A. It will suffice here to give the result obtained for the velocity potential $\phi$ in the immediate neighbourhood of the torus:
\bg
\phi_{\mathrm{torus}} = a^2\dot{R}\left[ -{x\over \rho^2}+{1\over 2 R} \log {8R\over \rho} +{1\over 2R} {x^2\over \rho^2}\right]+O(a^2\dot{R}/R^2).
\ela{finphi2}
\ee
Here the notation is essentially that used earlier in section 2.1 with $\rho^2 = x^2 + y^2$. Note that, relative to an observer moving with the torus, the normal velocity on $\rho=a$ is $\dot{a}$, as required. Also
\bg
\int_0^{2\pi} 2\pi a(R+a\cos\theta)\, {\ptl \phi\over\ptl \rho}\Big|_{\rho=a} \, d\theta = 0,
\ela{volc}
\ee
consistent with volume conservation. Moreover, if we wish to create a potential flow which conserves kinetic energy, with the cross sectional area decreasing as $R^{-3/4}$, we need only change the middle term of \er{finphi2} to $(3/4 R) \log  (8R/ \rho)$, with a 
corresponding addition of a multiple of \er{basepot} to the potential function. 

\begin{figure} % float placement: (h)ere, page (t)op, page (b)ottom, other (p)age
  \centering
  % file name: C:/Users/Steve/Desktop/RESEARCH/Newton2012/newtonpaper/tor1crop.eps
  \includegraphics[bb=0 0 370 196,width=3in,height=1.59in,keepaspectratio]{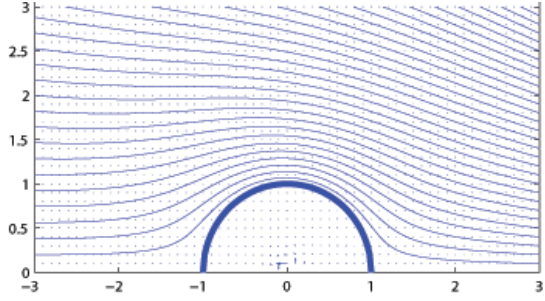}
  \caption{Flow lines for the velocity field with  potential given by \er{torp} with $1/R = 0.2$.}
  \label{fig:tor}
\end{figure}

Now we can obtain \er{finphi2} directly by observing that $\phi_0\equiv -a^2\dot{R}x/\rho^2$ is the perturbed potential for 2D flow past a cylinder. In 3D we need to solve
\bg
\phi_{xx}+\phi_{yy}= \lap\phi = -{1\over R+x} \, \phi_x,
\ee
and with $\phi = \phi_0+\phi_1+\dots$ we would have $\lap\phi_1=-R^{-1}\ptl\phi_0/\ptl x$.
Thus
\bg
\phi_1 = -{1\over 2R}\,  x\phi_0 
\ee
plus a harmonic function. The latter must be proportional to $\log\rho$ in order to satisfy (up to a function of time) \er{volc}, yielding \er{finphi2}. The terms involving $R^{-1}$
come from the shrinking of the cross-section as the torus expands, and the effect of curvature of the axis of the torus.
We show in figure \ref{fig:tor} a plot of the flow lines corresponding to the potential
\bg
\phi_{\mathrm{torus}}  =   -x -{x\over \rho^2}-{1\over 2 R}\,  \log \rho +{1\over 2R} \, {x^2\over \rho^2}+{1\over 4 R} \, (x^2+y^2),
\ela{torp}
\ee
where we have added a dilation to make the normal velocity vanish on $\rho=1$.

For the Sadovskii dipole {(not the snail, so dipole volume is preserved)} we can proceed similarly. Let $\phi_0$ be the dipole's 2D  exterior flow. Then
\bg
\phi  = \phi_0 -{1\over 2R}\, x\phi_0+\phi_1+\dots,
\ela{sadext}
\ee
where $\phi_1$ is a harmonic function, which sets the appropriate normal velocity at the boundary of the dipole; see below. In the following sub-section we match \er{sadext},  {modified to produce the snail}, to a potential flow defined on the scale of the toroidal dipole, using the functions developed in appendix A. 

\subsubsection{A uniformly valid solution at leading order}

Our aim now is to exhibit a uniformly valid and compatible dipole to leading order in the sense that the first-order terms necessary to describe the topology of the flow lines are included. We know that the zeroth-order dipole is the Sadovskii vortex, and what follows is an approximate treatment of the modifications which produce the snail. 

Let $\tilde{S}$ denote the cross-section of the dipole in $(\xi,\eta)$ coordinates. Referring back to the coordinates in section 2.1, we define $S$ by
\bg
(x,y)\in S \; \Leftrightarrow \; (\xi,\eta) \in \tilde{S}.
\ee
Then in the vicinity of the dipole, including both interior and exterior, the solution we seek has the form
\bg
\uv = -\dot{R}\iv + \uv_{\mathrm{dipole}} +\uv_{\mathrm{shrink}}.
\ela{leading}
\ee
The first two terms on the right 
of \er{leading} describe the instantaneous flow for the Sadovskii dipole relative to the co-moving coordinates, and in the exterior we include the correction in \er{sadext} associated with the ``squeeze'' flow. The area of the dipole cross-section is 
${\mathcal A}\equiv A(R_0/R)^{3/2}$.  {In the intermediate region} $\mathcal A \ll x^2+y^2\ll R^2$
we will have
\bg
\uv_{\mathrm{dipole}} \sim -{k_1\dot{R}\over \pi}{\mathcal A} \nabla\left[{x\over \rho^2}-{1\over 2R} {x^2\over\rho^2}\right],\; \rho^2=x^2+y^2,
\ee
for some positive constant $k_1$. 

The term $\uv_{\mathrm{shrink}}$ represents the potential flow due to dipole shrinkage.   Thus $\uv_{\mathrm{shrink}}= \nabla \phi_{\mathrm{shrink}}$ outside the dipole, where in the intermediate region we may write
\bg
\phi_{\mathrm{shrink}} =  -{k_2\over \pi}{\dot{R}\,  {\mathcal A}\over R} \, \log{ (8 R/\rho)} + \phi_{\mathrm{add}},
\ela{phi3phi1equation} 
\ee
where $k_2$ is a constant to be determined, and $\phi_{\mathrm{add}}$ is harmonic and $O(\rho^{-2})$ for large $\rho$. Note that within the dipole the shrinkage of the snail in the $(x,y,z)$ frame is a uniform contraction equivalent to that of a volume-preserving torus, but the boundary  of the dipole contracts faster owing to the stripping away of vorticity.

Other first-order terms are ignored. For example the {sink} distribution { contained in $\uv_{\mathrm{shrink}}$} is over $S$, instead of the perturbed vortex which includes the tail. There is thus a ``boundary-layer'' of vorticity missing here, associated with a tangential jump in velocity within first-order terms.  Also the tangential component of the  perturbation flow in the exterior has not been matched to an interior flow perturbation. To put this another way, \er{leading} captures the shrinking snail, but makes small errors in its shape.
 
Let us now consider the exterior potential flow relative to the fluid at infinity. Given the instantaneous centre curve of the Sadovskii vortex, we surround the structure by a concentric torus
of cross-sectional area large compared to the Sadovskii vortex area, but small compared to $R^2$. Call the surface of this torus $\ptl T$.
In the region outside of $\ptl T$ we shall represent the potential of the flow relative to the fluid at infinity  in the form
\bg
\phi_{\mathrm{ext}} = -2\dot{R} {\mathcal A} (k_1\Phi+k_3\phi/R),
\ee
where $\Phi$, $\phi$ are as given in appendix A, and $k_3$ is another constant to be determined. On $\ptl T$ we have, to leading order
\bg
\phi_{\mathrm{ext}}\sim {k_1\over\pi}\, \dot{R}{\mathcal A}\left[ -{x\over \rho^2}-{1\over 2 R} \log {8R\over \rho} +{1\over 2R}\,  {x^2\over \rho^2}\right]+ {k_3 \dot{R} {\mathcal A}\over \pi R} \log {8R\over \rho}.
\ela{phiext}
\ee
Using \er{phiext} we may compute the net flux of fluid into $T$, which must equal the rate of change of volume of the toroidal dipole (and the flux into the tail). We obtain
\bg
-4\pi\dot{R} {\mathcal A} (- k_1+k_3) = {d\over dt} \, 2\pi R {\mathcal A} = {d\over dt}\,  2\pi R A \left( \frac{R_0}{R}\right)^{3/2} = -\pi \dot{R} {\mathcal A}.
\ee
Thus 
\bg
k_{3} = \tfrac{1}{4}+k_1.
\ee
On the other hand, approaching $\ptl T$ from within we may write,
\bg
 \uv\sim \nabla \phi_{\mathrm{in}},
\ee
where
\bg
\phi_{\mathrm{in}} \sim {k_1\over\pi}\dot{R}\, {\mathcal A}\left[ -{x\over \rho^2}+{1\over 2R} {x^2\over \rho^2}\right]{-{\dot{R}{\mathcal A}\over \pi R}k_2 \;\log(8R/\rho)}.
\ela{phiin}
\ee
Comparing \er{phiext} and \er{phiin} we see that
\bg
{k_2}= \tfrac{1}{2} k_1-k_3,\quad
k_2=  { -\tfrac{1}{2}  k_1 -\tfrac{1}{4}} .
\ee
Here the term $-\tfrac{1}{2}  k_1$ contributes a correction to $\uv_{\mathrm{dipole}}$ to yield zero flux into the dipole, while the term $-\tfrac{1}{4}$ gives the flux into the dipole, flux  which is then expelled into the tail.

Since we have correctly established the flow through $\ptl T$ we are assured that the boundary condition in $\ptl S$ can be met for a suitable $\phi_{\mathrm{add}}$ {in \er{phi3phi1equation}}, so as to match the normal velocity associated with shrinkage due to geometry and erosion. 

%\subsubsection{Remark on the constancy of kinetic energy}
%
%Although we have argued that constancy  of kinetic energy implies a certain asymptotic dipole structure, we have not excluded the possibility that dipole kinetic energy is in fact
%{\em not} conserved. That is, can a finite kinetic energy be lost to the tail along with volume? We now consider this possibility.
%
%Again let $a\sim R^{-p}$, and  kinetic energy of order $R^{3-4p}$. The question is, can $p$ exceed 3/4? We saw in section 1 {\color{magenta}\emph{surely not section 1???}} that the flux of kinetic energy into the tail is of order  $R^{1-7p}$. 
%Let us determine kinetic energy as a function $F(R)$ which would yield this flux. Then
%\bg
%F^\prime (R) R^{1-p}\sim R^{1-7p},
%\ee
%so that $F = F_0+ f(R)$ where $F_0$ is a constant and $f\sim R^{1 -6p}$. But if $p>3/4$ we know that kinetic energy is decreasing as $R^{3-4p}$. Thus $F_0=0$ and $1-6p= 3-4p$, which is impossible if $p > 3/4$. Note that if $p=3/4$ a similar argument allows $F_0 > 0$ and $f\sim R^{-7/2}$.
%Thus we are forced to essentially maintain kinetic energy and $p=3/4$ stands as the exponent determining the only viable dipole structure.

\subsubsection{Summary}

We recapitulate the results of this section in anticipation of the redevelopment of the problem in the next section. We have established that the imposition of local conservation of energy leads to an eroding Sadovskii dipole. From this we deduce the existence of a tail to which the eroded vorticity is extruded. However  if correct this model should evolve naturally from the dynamics. In particular the scaling following from $p=3/4$ should evolve as an eroding structure with locally steady structure in the shrinking coordinates, and the flow of vorticity into the tail should be a derivable perturbation of the underlying Sadovskii eddy. It is just such a dynamical calculation that we now want to pursue.

This will entail a somewhat different approach in the coordinates used and the formulation of the underlying scaling of the dipole as an unknown. While this will involve a more systematic asymptotic theory, we will again encounter elements of the solution already exhibited. For example the terms $\xi\Omega_0$ and $-U_0$ on the right in \er{geneq1} embody the curvature of the cylindrical geometry, the former closely related to terms in section {4} indicated by the superscript `sq'', short for ``squeeze''. These terms arise from curvature of the vortex lines, a flow along the binormal which squeezes together curved anti-parallel vortex filaments and is a main cause of  shedding of vorticity into the tail.

\section{Full analysis of the perturbed Sadovskii dipole}

We therefore consider the structure of the perturbed dipole including the effect of shed vorticity. % and the calculation of the perturbed shape of the dipole boundary. 
This will demand a more elaborate formulation than we have adopted in section 3, for we now seek to determine dynamically how the perturbation of the Sadovskii dipole shape leads to erosion of vorticity and therefore determines the rate of shrinkage and thus the speed of the dipole. 
%This will entail computing the perturbed boundary flow of the dipole.
% which carries vorticity into the tail and modifies its shape. 
In effect the parameter $p$ in the discussions of section 2 is now an unknown to be determined from an asymptotic solution of Euler's equations valid for large $a/R$. We shall find that $p$ may be computed numerically from the condition that a scaling actually exists, i.e.\ that in suitable coordinates the structure appears steady, just as to leading order the snail is steady in local $(\xi,\eta)$ coordinates. We shall try to maintain a certain part of the notation of the previous sections. However there will be departures and we will work mainly in terms of a stream function, so the reader should regard this section as largely self-contained in its notation.

\subsection{Inner expansion about a steady Sadovskii vortex} 

We seek a solution for the vortex pair evolution at large radii $R(t)$ for which it is helpful to set $r \varpi = \omega = \partial u_z/ \partial r - \partial u_r / \partial z$ and to solve \er{exact2} using  a Stokes stream function $\psi$,  $u_r = - r^{-1} \partial \psi/\partial z$, $u_z =  r^{-1} \partial \psi/\partial r$. Introducing these into \er{exact2}, we seek to solve the vorticity equation
\begin{align}            
\frac{\partial \varpi}{\partial t}  & +  \frac{1}{r} \, \frac{\partial(\psi,\varpi)}{\partial(r,z)}  = 0 , \quad
%\label{eqvort}
\varpi  = \frac{1}{r}\,  \frac{\partial}{\partial r}  \left( \frac{1}{r} \, \frac{\partial\psi}{\partial r} \right) + \frac{1}{r^2} \, \frac{\partial^2\psi}{\partial z^2} \, . 
\label{eqlink} 
\end{align}
This simplifies a little if we replace the radial coordinate $r$ by $\tfrac{1}{2}r^2$ and this motivates the change of variables from $(r,z,t)$ to $(\xi, \eta, \tau)$ given by 
\begin{equation}
r^2 = R^2 + 2 a R \xi = R_0^2 ( g^2 + 2 f g \xi )  , \quad
z = a \eta = R_0 f \eta, \quad
d\tau / dt  = \omega_0  R /R_0= \omega_0 g , 
\label{eqcoordchange}
\end{equation}
with $R_0$ and $\omega_0$ dimensional reference quantities as before. We have introduced dimensionless radii  given by $a(t)= R_0 f(\tau)$ and $R(t) = R_0g(\tau) $. The transformation differs only in minor ways from that introduced earlier in section { 3.1}. For the fields we set 
\begin{align}
& \varpi(r,z,t) = \frac{\omega_0}{R_0}\,  \widetilde{\varpi}(\xi,\eta,\tau) , \quad 
\psi(r,z,t)  = {\omega_0 R_0^3}{f^{2} g^{2}} \,  \widetilde{\psi}(\xi,\eta,\tau) . 
\end{align}
%\\
%& (u,v)(r,z,t) = \omega_0 R_0 f^{-2} g^{-2} (U,V)(\xi,\eta,\tau) . 
%
Dropping any tildes leaves the vorticity equation and vorticity--stream function link as 
\begin{align}
\frac{\partial \varpi}{\partial\tau} & - \frac{\gdot}{f}\,  \frac{\partial\varpi}{\partial\xi}\, 
 -  \frac{\gdot}{g}  \, \xi \, \frac{\partial\varpi}{\partial\xi} 
 - \frac{\fdot}{f } \left( \xi \, \frac{\partial\varpi}{\partial\xi}+ \eta\,  \frac{\partial\varpi}{\partial\eta} \right) 
+   \Jc(\psi,\varpi) = 0 ,
\label{eqZvort}
\\
\varpi & =  \frac{\partial^2 \psi}{\partial\xi^2} + \frac{1}{1+2fg^{-1} \xi} \, \frac{\partial^2\psi}{\partial\eta^2} \, , 
\label{eqZlaplace} 
\end{align}
where we use $\Jc$ for a Jacobian with respect to the $(\xi,\eta)$ coordinates, and a dot (in this section only) for a $\tau$-derivative of $f$ or $g$. This formulation is exact but we have in mind $g =   R/R_0 \gg1$ and $f = a/R_0 \ll1$ for large times and that these are slowly varying, that is, 
\begin{equation}
f\ll1, \quad g\gg1, \quad \fdot/f \ll1, \quad \gdot/g \ll 1 ; 
\label{eqassume}
\end{equation}
these may be verified \emph{a posteriori}. We remark that the assumption $p=3/4$ in section~2 leads to $f\sim t^{-1}$,  $g\sim t^{4/3}$, so that, as before,  an expansion for large $t$ is implied. Our aim now is to obtain a value for $p$ asymptotically by analysis of the shedding of vorticity into the tail.

Thus for an \emph{inner solution}, that is valid for $(\xi,\eta)=O(1)$, we drop the $2fg^{-1} \xi$ term in (\ref{eqZlaplace}) and the \emph{frame contraction} terms involving $\fdot/f$ and $\gdot/g$ in (\ref{eqZvort}) at leading order. We use an inner expansion
\begin{equation}
\varpi = \varpi_0 + \frac{f_0}{g_0}\,  \varpi_1 + \cdots , \quad
\psi = \psi_0 +  \frac{f_0}{g_0}\,  \psi_1 + \cdots , \quad
g = g_0 + g_1 + \cdots , \quad
f = f_0 + f_1+ \cdots , 
\label{eqexp1} 
\end{equation}
in which we will find that the $\varpi_1$ and $\psi_1$ are of order unity. 
This gives, at leading order, equations for purely two-dimensional Euler flow, 
\begin{equation}
 \frac{\partial\varpi_0}{\partial\tau} - c_0 \, \frac{\partial\varpi_0}{\partial\xi} +  \Jc(\psi_0,\varpi_0) = 0 , \quad
\varpi_0 =  \frac{\partial^2\psi_0}{\partial \xi^2}+ \frac{\partial^2\psi_0}{\partial\eta^2} , 
\label{eqlovort}
\end{equation}
with $c_0$ defined by 
\begin{equation}
c_0 = \gdot_0 / f_0 . 
\label{eqc0def} 
\end{equation}
Now all we have done so far is valid for any $f_0(\tau)$ and $g_0(\tau)$ and so without further information $c_0$ could also depend on $\tau$. However we are seeking a leading order approximation as the steady Sadovskii vortex with continuous velocity. We thus set the vorticity, stream-function and speed, that is $(\varpi_0,\psi_0,c_0)$, to be one of the family of such vortices, with $c_0$ taken as constant. We will later choose one with $\varpi_0=\pm1$ in the two lobes, and $c_0=1$, but for the moment the choice is arbitrary. Thus, (\ref{eqc0def}) provides a single ODE linking $f_0$ and $g_0$; we need a further ODE to close the system, which will emerge at the next order. 

Although the choice of our leading order steady solution is arbitrary, the fact that it is one of a family has important implications. It means that an infinitesimal translation, 
\begin{equation}
\varpi_0^{\trans} =\frac{\partial\varpi_0}{\partial\xi} , \qquad
\psi_0^{\trans} = \frac{\partial\psi_0}{\partial\xi} ,
\label{eqzeta0transdef} 
\end{equation}
satisfies the linear equations  
\begin{equation}
 - c_0  \, \frac{\partial\varpi^{\trans}_0}{\partial\xi} +  \Jc(\psi^{\trans}_0,\varpi_0) + \Jc(\psi_0,\varpi^{\trans}_0) = 0 , \qquad
 \psi_0^{\trans} = \Gc \varpi_0^{\trans}. 
\label{eqzeta0transdefeq} 
\end{equation}
For a solution $(\varpi_0,\psi_0,c_0)$, a rescaled solution is 
$(\varpi_0(\lambda\xi,\lambda\eta), \lambda^{-2} \psi_0(\lambda\xi,\lambda\eta), \lambda^{-1}c_0)$  for any $\lambda$. Thus, taking the derivative with respect to $\lambda$ at $\lambda = 1$, we obtain a solution giving an infinitesimal change of scale
\begin{equation}
\varpi_0^{\scale} = \xi \, \frac{\partial\varpi_0}{\partial\xi} + \eta \, \frac{\partial\varpi_0}{\partial\eta}, \qquad
\psi_0^{\scale} = \xi  \, \frac{\partial\psi_0}{\partial\xi}  +  \eta \, \frac{\partial\psi_0}{\partial\eta} - 2 \psi_0 ,
\label{eqzeta0scaledef} 
\end{equation}
which obeys
\begin{equation}
 - c_0 \,  \frac{\partial \varpi^{\scale}_0 }{ \partial\xi} +   c_0  \varpi_0^{\trans} 
 +  \Jc(\psi^{\scale}_0,\varpi_0) + \Jc(\psi_0,\varpi^{\scale}_0) = 0 , \qquad
 \psi_0^{\scale} = \Gc \varpi_0^{\scale}. 
 \label{eqzeta0scalepde} 
\end{equation}
Here we have introduced $\Gc$ in (\ref{eqzeta0transdefeq}), (\ref{eqzeta0scalepde}) as the operator inverting the Laplacian in (\ref{eqlovort}), that is integration against the kernel 
\begin{equation}
G(\xi,\eta) = (4\pi)^{-1} \log ( \xi^2 + \eta^2) 
\label{eqGkernel} 
\end{equation}
in infinite $(\xi,\eta)$ space. 

%{\bf Will need to align labels with elsewhere, e.g., (46).}

Having dealt with the leading order problem we now write down the first order equation, in which the neglected terms involving $\fdot/f$, $\gdot/g$ in (\ref{eqZvort}) and $2fg^{-2}\xi$ in (\ref{eqZlaplace}) are reintroduced to drive corrections $(\varpi_1,\psi_1)$ to the fields:
\begin{equation}
\frac{\partial\varpi_1}{\partial\tau} - c_0  \frac{\partial\varpi_1}{\partial\xi} 
 - c_1 \frac{\partial\varpi_0}{\partial\xi} 
-  c_0  \xi \frac{\partial\varpi_0}{\partial\xi} + c_0 p_1 \left(\xi \frac{\partial\varpi_0}{\partial\xi}  + \eta\frac{\partial\varpi_0}{\partial \eta} \right)
+  \Jc(\psi_0,\varpi_1) + \Jc(\psi_1,\varpi_0)  = 0 , 
\label{eqvort1}
\end{equation}
\begin{equation}
\varpi_1 =  \frac{\partial^2 \psi_1}{\partial\xi^2} + \frac{\partial^2 \psi_1}{\partial\eta^2} - 2 \xi  \, \frac{\partial^2\psi_0}{\partial \eta^2}\,   . 
\label{eqvortsf1}
\end{equation}
Here we have made use of  (\ref{eqassume}) and (\ref{eqc0def}), and defined
\begin{equation}
c_1 = \frac{g_0}{f_0^2} \, ( \gdot_1 - c_0 f_1) , \quad
p_1 = - \frac{\fdot_0 g_0 }{c_0f_0^{2}} \, .  
\label{eqc1d1def} 
\end{equation}
The quantity  $p_1$ will turn out to be the same as the exponent $p$ in section 2. 
To deal with this first order problem, first invert (\ref{eqvortsf1}) as 
\begin{equation}
\psi_1 = \Gc \varpi_1 + \psi_1^{\squeeze} , \quad
\psi_1^{\squeeze}  \equiv  2  \Gc\left(\xi  \frac{\partial^2\psi_0}{\partial\eta^2}\right) 
\label{eqsqueezepsi} 
\end{equation}

and then use (\ref{eqzeta0transdef}, \ref{eqzeta0scaledef}) to rearrange (\ref{eqvort1}) as 
\begin{align}
\frac{\partial\varpi_1}{\partial\tau} - c_0  \frac{\partial\varpi_1}{\partial\xi} +  \Jc(\psi_0,\varpi_1) 
& = 
  c_0  \xi \frac{\partial\varpi_0}{\partial\xi}  - \Jc(\Gc \varpi_1 ,\varpi_0) - \Jc(\psi_1^{\squeeze},\varpi_0) 
  \notag \\
 & +  c_1 \varpi_0^{\trans} 
 -  c_0 p_1 \varpi_0^{\scale} 
 \label{eqfullzeta1}
 \end{align}
On the left-hand side we have advection of vorticity $\varpi_1$ in the basic flow field of the Sadovskii vortex; on the right-hand side are the remaining terms. Several remarks are in order. This equation is ``driven'' by the terms $c_0 \xi  {\partial\varpi_0}/{\partial\xi}$ and  $\Jc(\psi_1^{\squeeze},\varpi_0)$, in that if these terms were absent a solution would be $\varpi_1=0$, $c_1=p_1=0$. These driving terms have an amplitude that is independent of $\tau$ thanks to our scalings, and so are fixed, constant in time. Although the driving terms are constant, the solution $\varpi_1(\xi,\eta,\tau)$ need not be steady and in fact will generally not be, as it will acquire pieces of $\varpi_0^{\trans}$ corresponding to drift in the $\xi$-direction, and $\varpi_0^{\scale}$ corresponding to a change in scale; see (\ref{eqzeta0transdef}--\ref{eqzeta0scalepde}). On the other hand we can eliminate these terms by suitable choice of $c_1$ and $p_1$ --- we will check this numerically in due course --- and with this choice we expect to be able to obtain a solution $\varpi_1$ independent of $\tau$. Note that this choice is available to us as originally the functions $f$ and $g$ were arbitrary rescalings --- we can choose to fix them order by order. This imposition of a solvability condition, that the first order solution remains bounded uniformly in time, gives a solution representing the modified Sadovskii vortex, traveling outwards according to $g_0(\tau)$ and shrinking through shedding vorticity according to $f_0(\tau)$. 

So, we suppose we have converged to a steady solution $\varpi_1(\xi,\eta)$ with constants $c_1$ (which is not of use to us as it involves $f_1$ as a new unknown) and $p_1$ which gives a second ODE linking $f_0$ and $g_0$ in (\ref{eqc1d1def}). Together with (\ref{eqcoordchange}), (\ref{eqc0def}) we obtain 
\begin{equation}
f_0 \propto \tau^{p_1/(1 + p_1)} \propto (\omega_0  t)^{-1} , \quad
g_0 \propto \tau^{ 1/(1 + p_1)} \propto (\omega_0 t)^{1/p_1}, \quad
\tau \propto (\omega_0 t)^{1+ 1/p_1} , 
\end{equation}
and we anticipate $p_1>0$ so that $g_0$ increases with $t$ and the approximations are all self-consistent. Here we may identify $p_1=p$, the exponent introduced in section 2.1.

\noindent

Before we set about solving to find $p_1$ we again comment that the term in $\psi^{\squeeze}$ corresponds to the leading order effect of curved vortex lines creating a flow that drives the two lobes of the Sadovskii vortex together, a weak but controlling effect in our expansion. Together with this term, we should note that when we invert minus the Laplacian and write down $\psi_0 = \Gc \varpi_0$ and $\Gc\varpi_1$, for example in (\ref{eqsqueezepsi})  we could add on a component which is harmonic in the $(\xi,\eta)$ plane. In fact fundamentally this is how the distant structure of the vortex would feed into the inner solution, modifying vortex shape and motion. It is clear that the terms that would be incorporated would take the form of a multipole expansion: the first would appear at the level of $\Gc\varpi_1$ and would correspond to uniform flow. This could be absorbed into $c_1$ (a Galilean transformation) but would not affect the vortex structure or $p_1$. If we went to the level of $\Gc\varpi_2$ it would be necessary to introduce an external strain field which would have an effect, but fortunately this is beyond the order we need. 

\noindent
%{\bf Steve: I could flesh this out more, but more importantly it would be good to link this to your discussion of the external potential field as it is a link that is clearly worth making. Fleshing it out more means estimating the size of the corrections. } 

\subsection{Formulation in terms of contours and numerical solution.} 

Now the above is written as if the vorticity fields are smooth, but in fact we are working about the Sadovskii vortex in which the vorticity field is piecewise constant, and so to actually solve the above problem we need to work, not with the fields $\varpi_0$, $\varpi_1$, but instead using contour dynamics. We have in mind here the asymptotic state of the dipole pair for large radius $R$, where erosion has led to the vorticity being effectively constant in each lobe, and the eroded edge is taken as a discontinuity. We thus need to further manipulate the equations, working in the $(\xi,\eta)$ plane with the use of polar coordinates $(\rho, \theta)$ in this plane when needed. To commence, note that although equation (\ref{eqfullzeta1}) is complicated, it does take the form 
\begin{equation}
\frac{\partial\varpi_1}{\partial\tau} + \Uv_0 \cdot \nabla  \varpi_1 + \Uv_1 \cdot \nabla \varpi_0 = 0 
\label{eqvort2}
\end{equation}
(we will express $\Uv_0$ and $\Uv_1$ explicitly below), which is the linear piece of the full equation 
\begin{equation}
\frac{D\varpi}{D\tau} \equiv  \frac{\partial\varpi}{\partial\tau} + \Uv \cdot \nabla  \varpi = 0, \quad
 \varpi = \varpi_0 + \varpi_1 + \cdots, \quad
 \Uv = \Uv_0 + \Uv_1 + \cdots. 
\label{eqvort3}
\end{equation}
The leading piece of this is $\Uv_0\cdot\nabla\varpi_0=0$
and gives the steady Sadovskii vortex with the velocity $\Uv_0 = (U_0,V_0)$ and vorticity linked to the total stream function $\Psi_0 = \psi_0 + c_0 \eta$ (including the flow past the vortex) via
\begin{equation}
U_0 = - \frac{\partial\Psi_0}{\partial\eta}, \quad
V_0 = \frac{\partial\Psi_0}{\partial\xi}, \quad
\varpi_0 =\frac{\partial V_0}{\partial\xi} -  \frac{\partial U_0}{\partial\eta} . 
\end{equation}

\begin{figure} % float placement: (h)ere, page (t)op, page (b)ottom, other (p)age
  \centering
  % file name: C:/Users/Steve/Desktop/RESEARCH/Newton2012/newtonpaper/PerturbedDipole.eps
  \includegraphics[bb=0 0 493 296,width=4in,height=2.4in,keepaspectratio]{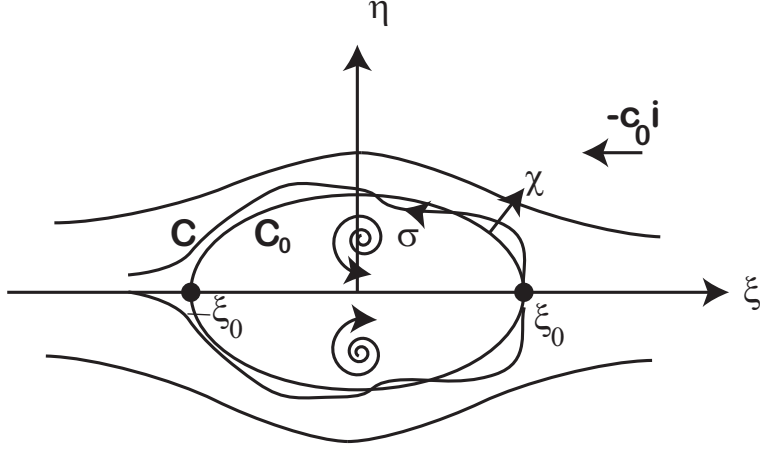}
  \caption{Schematic of the perturbed Sadovskii vortex.}

  \label{fig:PerturbedDipole}
\end{figure}

The vortex has vorticity $\varpi_0 = 1$ in a region bounded by the $\xi$-axis and a contour $C_0$ in the half-plane $\eta>0$ and $\varpi_0= - 1$ in the mirror image region; see 
figure~\ref{fig:PerturbedDipole}. Using the divergence theorem the corresponding stream function $\psi_0 = \Gc \varpi_0$ can be obtained by integrating over the boundaries and gives 
\begin{align}
\psi_0(\xi, \eta)    
 =  \frac{1}{4\pi} \int_{C_0}  &  \left\{
 \log |(\xi',\eta') - (\xi,\eta) | \, [ (\xi',\eta') - (\xi.\eta) ] \cdot (d\eta', - d\xi') 
\right.  
\notag \\
&  \left. 
 +  \log |(\xi',- \eta') - (\xi,\eta) |\,  [ (\xi',- \eta') - (\xi,\eta)] \cdot (-d\eta', - d\xi')  
\right\}
\notag\\
 & + \frac{\eta}{2\pi} \Bigl[ \xi' \log\sqrt{\xi'^2 + \eta^2} +  \eta \tan^{-1} (\xi'/\eta) - \xi' \Bigr]^{ \xi' = \xi_0 - \xi}_{ \xi' = -\xi_0 - \xi} 
\ela{eqSadsf} 
\end{align}
with the latter term giving the contribution from the integral along the base, that is the piece $-\xi_0 \leq \xi\leq \xi_0$ of the $\xi$ axis. For this to represent a steady vortex dipole embedded in a flow $(-c_0,0)$ at infinity we need the total stream function $\Psi_0 = \psi_0 + c_0 \eta$ to be zero on the contour $C_0$. This condition enables $C_0$ to be found for a given $c_0$, for example using a collocation method as described in \cite{ST}.\footnote{In equation \er{eqSadsf} we correct a misprint in \cite{ST}, noting that their stream function is taken with the opposite sign to ours. 

}

With  $C_0$ known at least numerically, we can use a coordinate system based on arc-length $\sigma$ along the contour and a coordinate $\chi$ that measures distance perpendicular to the contour $C_0$; see figure \ref{fig:PerturbedDipole}. 
The corresponding metric is then 
\begin{equation}
ds^2 =   d\chi^2 + h^2 d\sigma^2 , \quad
h = 1 + \kappa(\sigma) \chi, 
\end{equation}
where $\kappa$ is the curvature of the curve $C_0$ at the point given by $\sigma$.
We need to recast the first equation of (\ref{eqvort3}) in a contour dynamics setting. We take the non-zero constant vorticity to be $\varpi=1$ in the upper half-plane, confined by a time-dependent contour which we call $C$ and suppose (with mild abuse of notation) given by a function $\chi = C(\sigma,\tau)$ (see figure \ref{fig:PerturbedDipole}). The situation in the lower half plane is mirror symmetric. Now $C$ is a material curve and so 
\begin{equation}
\frac{D}{D\tau} \, ( C(\sigma,t)- \chi ) = 0 , 
\end{equation}
or
\begin{equation}
 \frac{\partial C}{\partial\tau}  = \Uv \cdot \left(\nv -   \frac{1}{h}  \,  \frac{\partial C}{\partial\sigma} \, \tv\right)  \, \, 
 \Big|_{\chi = C(\sigma,\tau)}  \,  , 
 %\quad h = 1+\kappa(\sigma)\chi 
\end{equation}
with $\tv = h \nabla\sigma$ and $\nv = \nabla\chi$ being tangential and normal unit vectors. %and $\kappa(\sigma)$ the curvature at each point on $C_0$. 
The unperturbed problem has $C(\sigma,\tau) = C_0(\sigma,\tau) \equiv 0$ and $\Uv_0\cdot\nv =0 $. At the first order we set $C (\sigma,\tau)= C_1(\sigma,\tau) + \cdots$, $\Uv = \Uv_0 + \Uv_1 + \cdots$ to obtain in the linear approximation, 
\begin{equation}
\frac{\partial C_1}{\partial \tau} =\left(  C_1 \,\frac{\partial\Uv_0}{\partial\chi} + \Uv_1 \right) \cdot\nv - \frac{\partial C_1}{\partial \sigma} \, \Uv_0 \cdot \tv. 
\end{equation}
evaluated on the curve $C_0$ given by $\chi=0$. With the use of the stream function we write 
\begin{equation}
\Uv_0 = - \frac{1}{h}\,  \frac{\partial \Psi_0}{\partial\sigma} \,  \nv 
                   + \frac{\partial \Psi_0}{\partial \chi} \, \tv
\end{equation}
and after a short calculation obtain 
\begin{equation}
\frac{\partial C_1}{\partial\tau} + \frac {\partial}{\partial\sigma} \left( C_1 \,  \frac{\partial \Psi_0} {\partial\chi} \right)  = \Uv_1\cdot\nv 
\label{eqC1pde}
\end{equation}
again evaluated on $C_0$. 
%{\bf Note I had to go around the houses to get (85), for example the unit vectors are independent of $\chi$... I have this written out elsewhere (in the old notation) and will check it again. If you want me to put more in I can. Annoyingly the final result is quite simple and I suspect there are easier ways to get there.} 
Setting 
\begin{equation}
\Phi_1 = C_1 \, \frac{\partial\Psi_0}{\partial\chi} \equiv C_1 \Uv_0 \cdot \tv
\end{equation}
we can write the equation in perhaps the most intuitive form
\begin{equation}
\frac{\partial \Phi_1}{\partial\tau} + \Uv_0\cdot\tv \, \frac {\partial \Phi_1}{\partial\sigma}   = (\Uv_0\cdot\tv)(\Uv_1\cdot\nv )
\label{eqfinalpde}
\end{equation}
%
%{\bf Steve: I'm running out of letters here and am not sure that $C_1$ and $\Phi_1$ represent the best notation. I did use $X_1$ and $Y_1$ in earlier notes but these letters are already in use with other connotations.} 
This is intuitive in that it represents advection of vorticity flux $\Phi_1(\sigma,\tau)$ between curves $C_0$ and $C_1$ along the unperturbed curve $C_0$, with a source term that involves the perpendicular component of the perturbation velocity $\Uv_1$. Note that as we approach the trailing stagnation point, where vorticity will peel off into the flow, $\Uv_0\cdot\tv\to0$ and so both the source term on the right-hand side is suppressed, and the quantity $\Phi_1$ will be seen to converge, even though $C_1$ must diverge there. 

\begin{figure} % float placement: (h)ere, page (t)op, page (b)ottom, other (p)age
  \centering
  % file name: C:/Users/STEVE/Desktop/Newton2012/newtonpaper/asympsingfig1.eps
%% 
{(a)}
\includegraphics[scale=0.32]{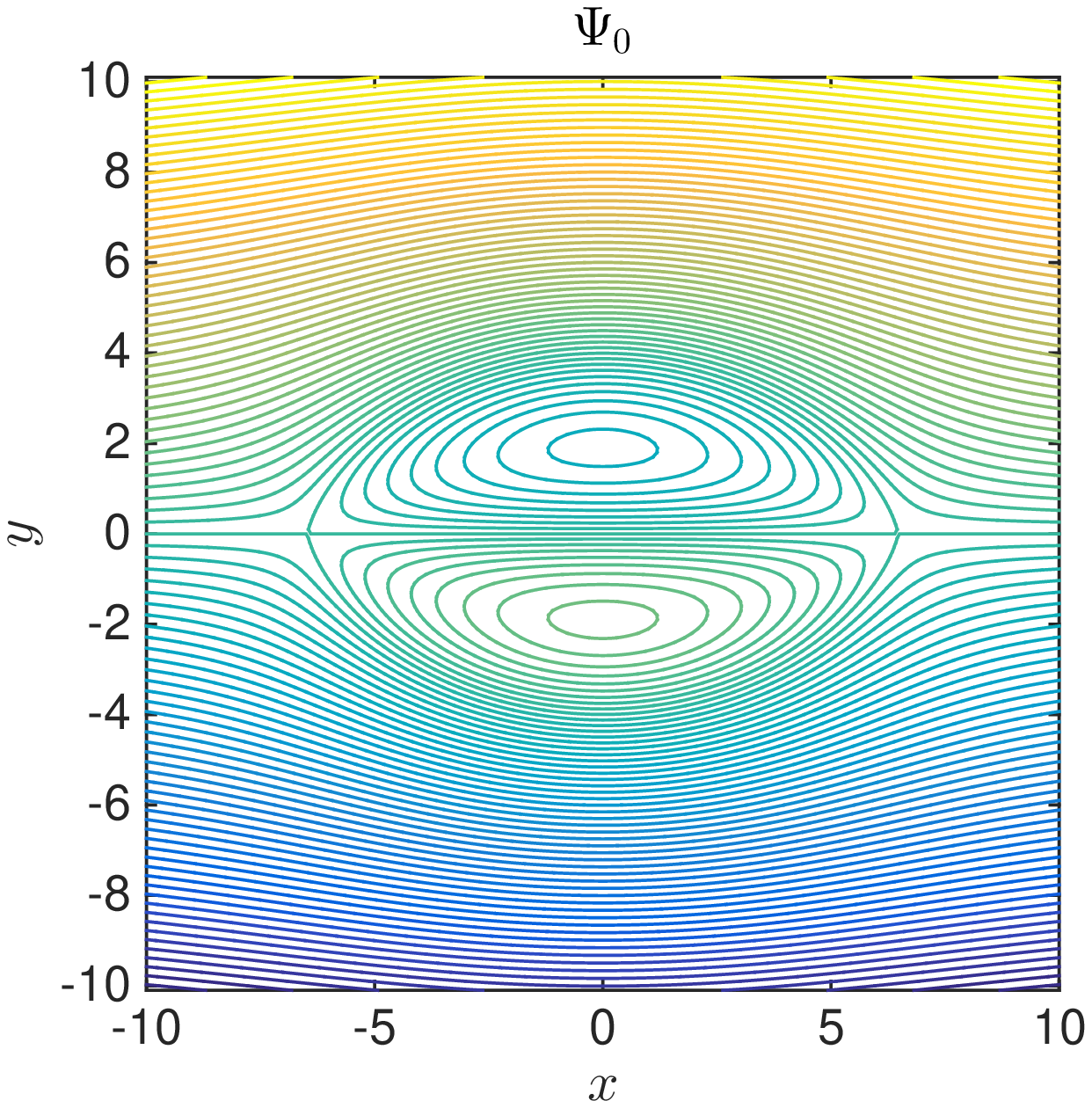}
{(b)}
\includegraphics[scale=0.32]{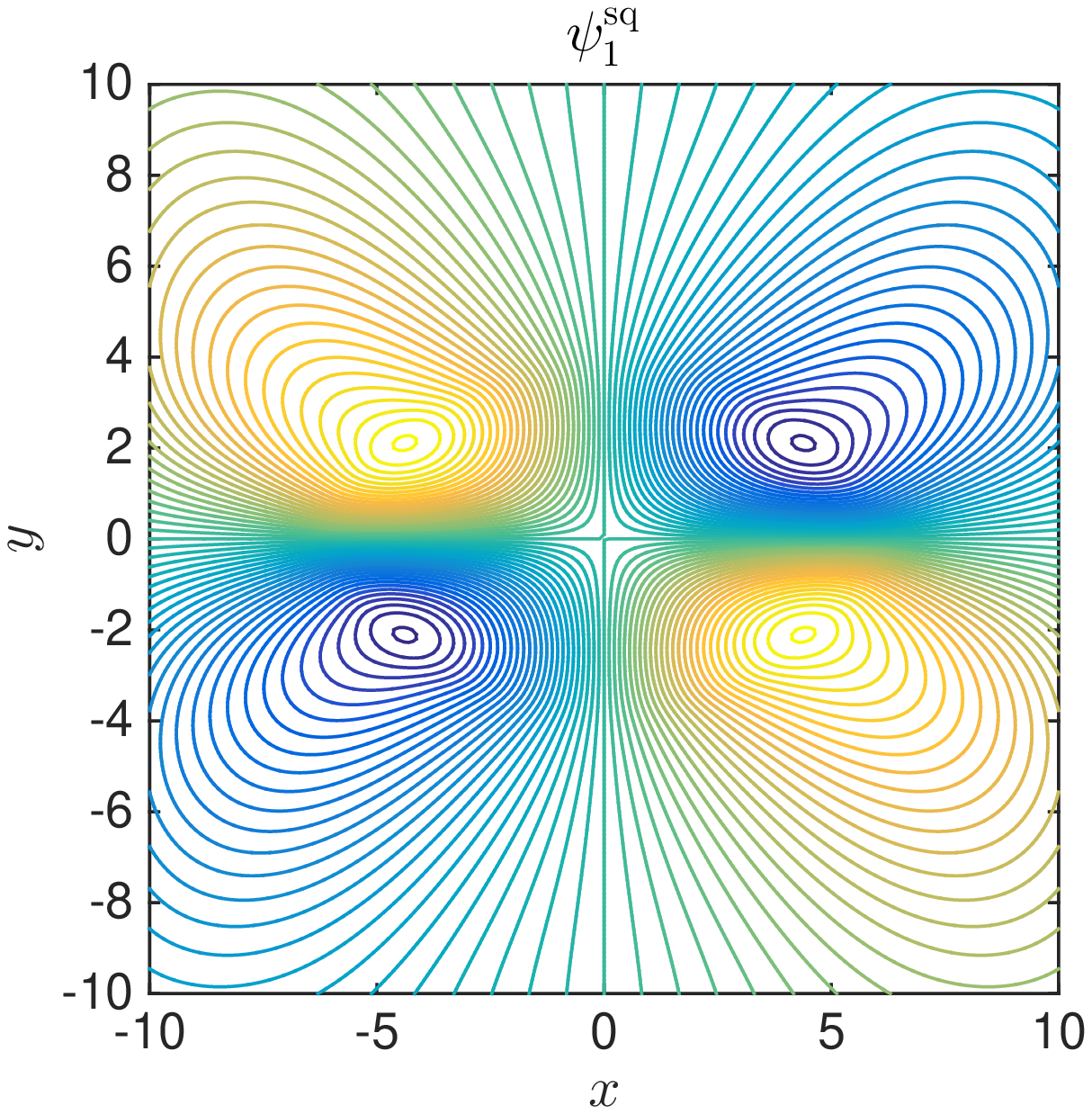}
  \caption{Flow field for (a) the Sadovskii vortex, given by  curves of constant $\Psi_0$, and (b) the flow driven from $\psi_1^{\squeeze}$, of roughly stagnation point form.}
  \label{fig:psisqueeze}
\end{figure}

With the key machinery in place, we indicate the numerical solution that aims to fix $p_1$, through time stepping the PDE (\ref{eqfinalpde}) until it can be made to converge to a steady state. Before we time step we evaluate the boundary of the Sadovskii vortex from \er{eqSadsf} following Saffman \& Tanveer and express this as a curve $\rho = \rho_0(\theta)$ in polar coordinates in the $(\xi,\eta)$-plane; the resulting flow field is depicted in figure \ref{fig:psisqueeze}(a). From this we may evaluate $\tv$ and $\nv$ along $C_0$ relative to polar coordinates. Then, for the left-hand side of (\ref{eqfinalpde}) we need $\Uv_0\cdot\tv$ which is obtained from the Sadovskii stream function in \er{eqSadsf} with $\Psi_0 = \psi_0 + c_0 \eta$ by finite differencing of $\psi_0$ as obtained numerically. Turning to the right-hand side of  (\ref{eqfinalpde}), $\Uv_1$ contains several components from (\ref{eqfullzeta1}), in order, 
\begin{equation}
\Uv_1 = \Uv_1^{\contraction} + \Uv_1^{\Gc} + \Uv_1^{\squeeze} +c_1  \Uv_1^{\trans} - c_0 p_1 \Uv_1^{\scale} . 
\end{equation}
The most straightforward of these are expressed in polar coordinates as 
\begin{align}
& \Uv_1^{\contraction}  = -  c_0 \rho \cos\theta (\cos\theta\, \rhovh - \sin\theta\, \thetavh) , 
\\
& \Uv_1^{\squeeze}  = - \frac{1}{\rho}\,  \frac{\partial\psi_1}{\partial\theta}^{\!\!\!\!\!\!\squeeze} \,  \rhovh + \frac{\partial  \psi_1}{\partial \rho}^{\!\!\!\!\!\!\squeeze} \,  \thetavh , 
\\
& \Uv_1^{\trans}  = - \cos\theta\, \rhovh + \sin\theta\, \thetavh , 
 \qquad
\Uv_1^{\scale}  = - \rho \rhovh . 
\end{align}
The term arising from vortex line curvature is $\psi_1^{\squeeze}$ which is a fixed flow field that can be evaluated once at the start of the computation. This is done rather crudely by evaluating $\partial^2 \psi_0/\partial\eta^2$ using finite differences, then applying $\Gc$ by approximating the integral as a finite sum over grid points, and finally finite differencing again. Streamlines of the resulting flow field are shown in figure \ref{fig:psisqueeze}(b); this has an approximate stagnation point form, pressing the two lobes of the vortex together as expected. 

Finally as we time step the PDE  (\ref{eqfinalpde}) the only term that cannot be pre-calculated is the feedback $\Uv_1^{\Gc}$ which is the flow arising from $\Gc \varpi_1$, from the perturbed contour and a functional of $C_1(\sigma, \tau)$. Now the unperturbed contour is $\rho = \rho_0(\theta)$ in polar coordinates, and the gap between this and the perturbed contour,  $\chi = C_1(\sigma,\tau)$, gives essentially a vortex sheet which has to be integrated as in (\ref{eqGkernel}) to obtain the corresponding flow. A short calculation shows that at a point $(\rho_0(\theta),\theta)$ on the contour the normal component that we need may be written as an integral over the contour, in terms of the dummy variable $\theta'$, 
\begin{align}
 \Uv_1^{\Gc} \cdot\nv & = 
\frac{1}{2\pi} \int_{0}^{\pi}  
\left\{ \frac{ \rho_0 \rho'_0 \sin (\theta-\theta') + (\partial_\theta \rho_0) [\rho_0 - \rho'_0 \cos (\theta-\theta')]}{ (\rho_0-\rho_0') ^2 + 4 \rho_0\rho'_0 
 \sin^2  \tfrac{1}{2} (\theta-\theta') } 
\,   \frac{j}{j'}\,  C'_1 
 - \tfrac{1}{2}  \cot \tfrac{1}{2}(\theta-\theta') C_1   
\right\} \, d\theta'
   \notag \\
  & -  \frac{1}{2\pi} \int_{0}^{\pi}  
\left\{ \frac{ \rho_0 \rho_0'  \sin (\theta+\theta') + (\partial_\theta \rho_0) [\rho_0 - \rho'_0 \cos (\theta+\theta')]}{ (\rho_0-\rho_0') ^2 + 4 \rho_0\rho'_0 
 \sin^2  \tfrac{1}{2}  (\theta+\theta') }
  \,   \frac{j}{j'} \, C'_1  
   - \tfrac{1}{2}   \cot \tfrac{1}{2}(\theta+ \theta') C_1   
   \right\} \, d\theta' 
   \notag\\
   &  +\pi^{-1} \log (\tan\tfrac{1}{2}\theta ) \,  C_1  , 
   \label{eqC1invert} 
\end{align}
where $\partial_\theta \rho_0= d\rho_0 /d\theta$, a prime denotes evaluation with respect to the dummy variable $\theta'$ and the Jacobian is given by 
\begin{equation}
j(\theta) ^{-1}\equiv \frac{d\sigma}{d\theta} = \left[ \left( \frac{d \rho_0}{d\theta}\right)^2 + \rho_0^2 \, \right]^{1/2} . 
\end{equation}
%

%{\bf Now again it is a bit messy to get (92), as per the comments below. I have some of this written in my old notes, but even these are sketchy. I will go through this again and see if either we should say more, or even if (92) is the best form in which to write it. We could write as an integral by $d\sigma'$ but then in actual practice it is done as a $\theta'$ integration. Anyway, some of these messier pieces are up for grabs --- feel free if you think we need more, or less here. I am also happy to put material in, that makes sure we both know what we are doing, that would ultimately be removed (I have much of it elsewhere, albeit in different notation.
%
%It is a bit hard to know what to include, this being a small component of an epic study!} 

%{\bf Note on figure \ref{fig:psisqueeze}(a): you may want to put this earlier somewhere! As the vortex is key to the whole study... I just put it with (b) for simplicity. } 

With this in place, we time-step the PDE (\ref{eqfinalpde}) in terms of $\Phi_1(\sigma,\tau)$, by evaluating $\Uv_1^{\Gc}\cdot\nv$ at each time $\tau$ and looking up all the other components of the flow field. We need to allow $c_1$ and $p_1$ to converge so that $\Phi_1(\sigma,\tau)$ becomes steady as $\tau\to\infty$, thus avoiding secular behavior. Essentially we have freedom about how this is done: any two conditions that fix the scale and the $\xi$-location of the vortex will suffice. We choose to require 
\begin{equation}
\Uv_1\cdot\nv = 0 \quad\text{ at }\quad \theta = 0 ,  \;  \pi/2 
\label{eqfixscale} 
\end{equation}
and so once all the components of $\Uv_1\cdot\nv$ are found the calculation of $c_1$ and $p_1$ is straightforward. 

\begin{figure} % float placement: (h)ere, page (t)op, page (b)ottom, other (p)age
  \centering
  % file name: C:/Users/STEVE/Desktop/Newton2012/newtonpaper/asympsingfig1.eps
%% 
(a)
\includegraphics[scale=0.3]{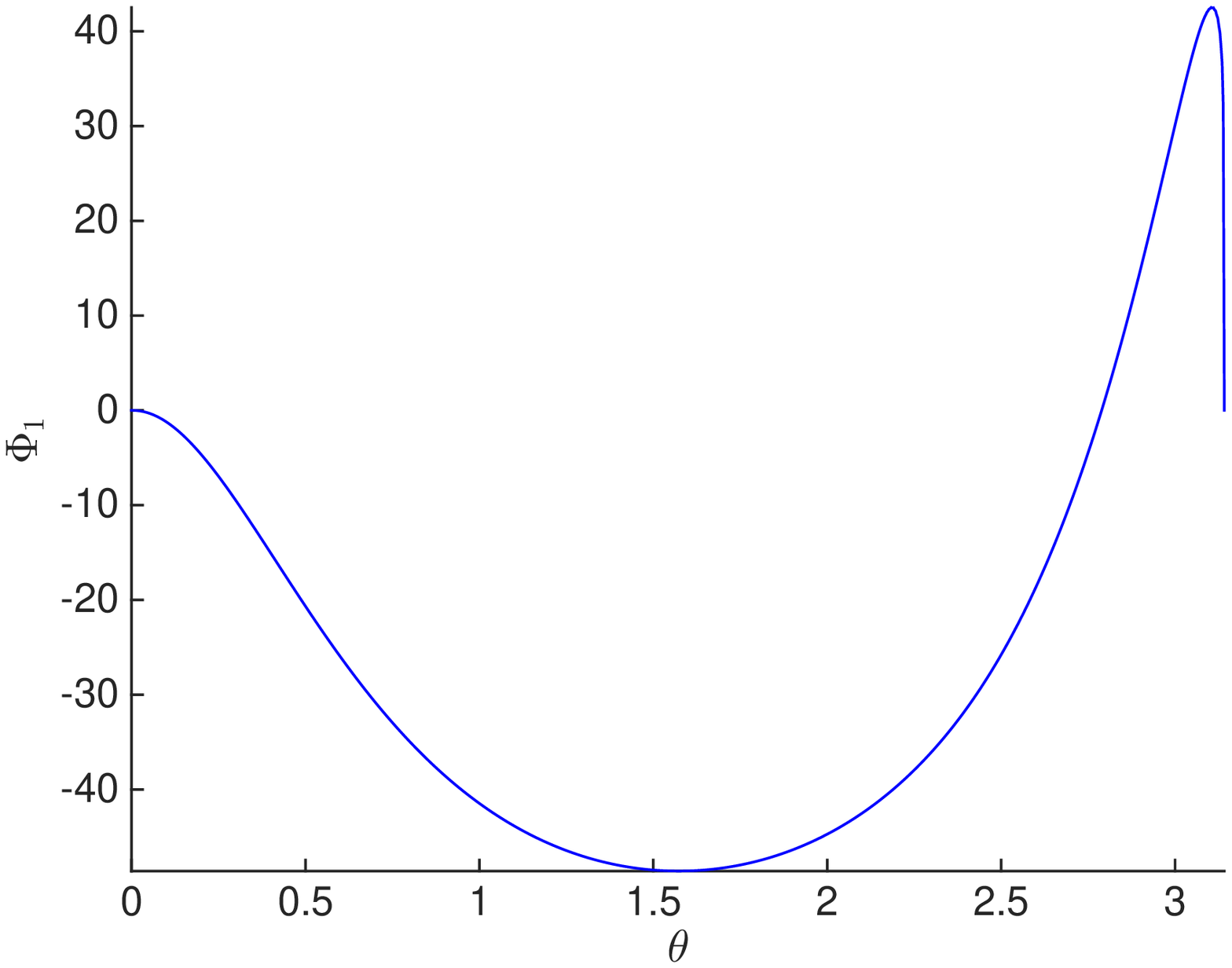}
(b)
\includegraphics[scale=0.3]{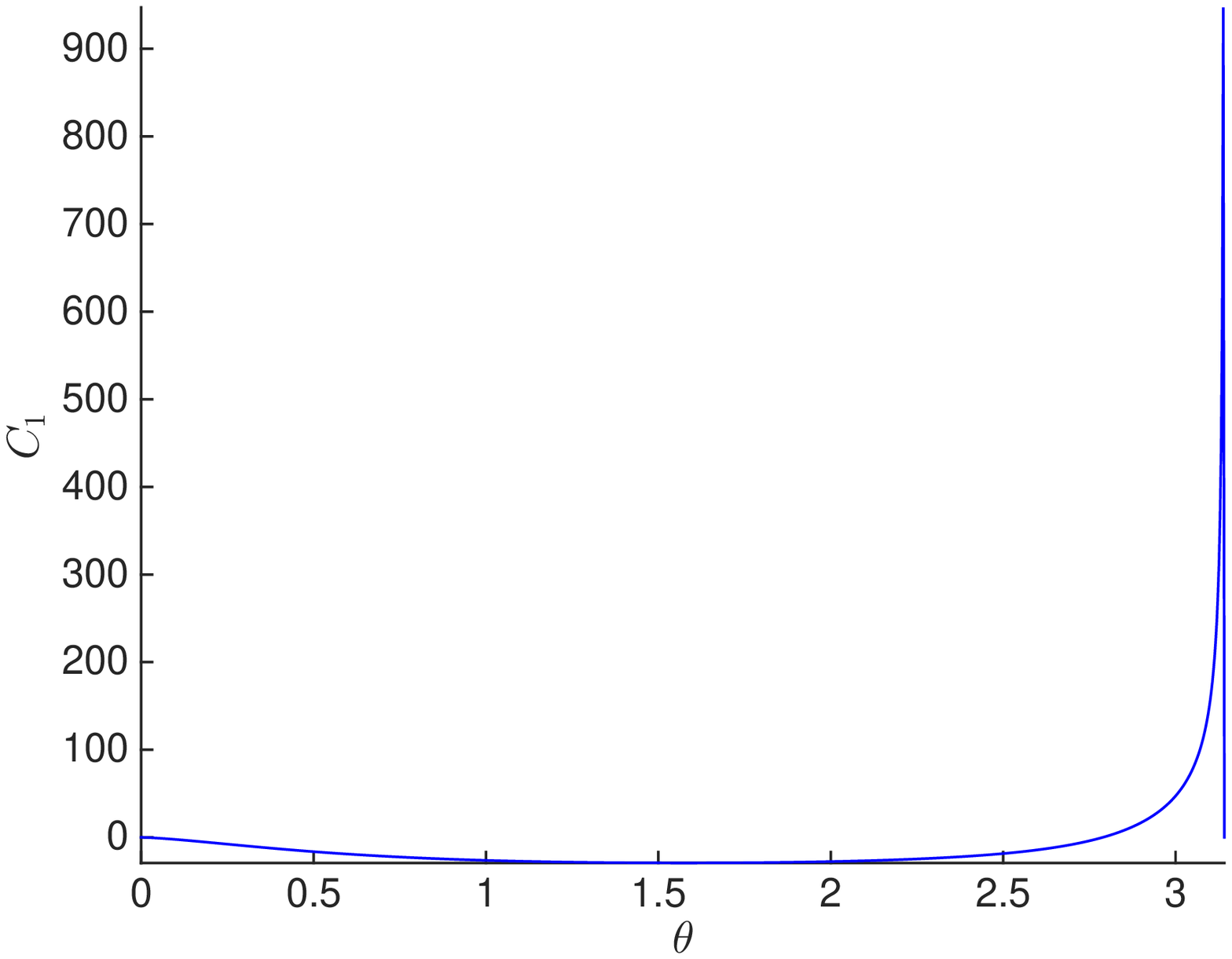}
  \caption{Steady correction to the Sadovskii vortex: (a) $\Phi_1$ and (b) $C_1$, valid for $\tau\to\infty$, are shown as functions of $\theta$.}
  \label{fig:Phi1C1}
\end{figure}

The result of time stepping is that $\Phi_1$ converges to a $\tau$-independent profile, depicted in figure \ref{fig:Phi1C1} with $p_1 \simeq 0.74$. This is in line with the theoretical value $p_1 = 3/4$ needed for energy conservation. Note that of the two driving terms, with only $\Uv_1^{\squeeze}$ we obtain $ p _1 = 0.37$, and with only $\Uv_1^{\contraction} $, $p_1 = 0.35$, so these each account for about half the effect. 

%\emph{\color{red}Whereas $p_1$ is robust at around $3/4$ when I vary resolution, etc., I get several values for $c_1$, for reasons I am not sure about... Of course this is not the important quantity, but I will think about this a bit more. Maybe I should be worrying a bit more about the structure of the solution near the rear stagnation point. But overall all looks fine!}

%{\bf Steve: note that  the pulling down of the curve in figure 7(a) may be an artifact. I will try to get these figures a bit more carefully. The fact tat both in (a) and (b) the curves go to zero is just my putting zero in the last point for various other reasons. (b) shows the tail as ejected.} 

%{\bf Steve: note that if PV's code brings in just one of these effects (I believe it neglects the squeeze effect, this suggests $g \propto t^{8/3}$ which is very big! Whether we could get this I'm not so sure though - I'm not very clear precisely what is in his formulation and what isn't. Also my results in this case are markedly different --- maybe the squeezing effect is actually really important in getting a tail to form... Needs further thought but this may all be something of a red herring...} 
%
%{\bf Note: I could add lots more figures --- I don't think we should, but if you feel something is missing let me know and I can provide it. I have plotted a lot more quantities in my attempt to get something which hangs together numerically.} 

Finally we remark on the formula (\ref{eqC1invert}): the feedback on the flow because the contour $C$ differs a little from $C_0$ amounts to calculating the flow from a vortex sheet of strength $C_1(\sigma,\tau)$ along the curve $C_0$. Such an integral has to be taken as a principal value, and here we have done this by removing explicitly the singular components from the integrands in (\ref{eqC1invert}), which are then placed in the final $\log (\tan\tfrac{1}{2}\theta )$ term. Taking the principal value  is appropriate as at a point $\sigma$ on the vortex sheet/thin layer the transverse flows across generated by the vorticity for $\sigma'>\sigma$ and $\sigma'<\sigma$ locally cancel out. However at $\theta=0$ and $\theta=\pi$, this argument fails: the vortex sheet comes to an abrupt end and in fact changes sign. (The curvature singularities and singular flow field here in the underlying Sadovskii vortex are explained in depth in \cite{ST}). This explains the presence of the logarithmic singularity at $\theta=0$, $\pi$ in the term $\log (\tan\tfrac{1}{2}\theta)$. Now in our calculations we have taken $\Uv_1\cdot\nv = 0$ at $\theta=0$ in (\ref{eqfixscale}) which keeps $C_1=0$ there (see (\ref{eqC1pde})) and removes immediate difficulties with this term. For $\theta=\pi$ the singular term is present, and is part of the flow field that leads to the ejection of vorticity from the rear of the vortex pair. 

\section{Numerical simulation of the snail}
We complement the analysis offered in previous sections with the results of a direct numerical simulation of the evolving axisymmetric vortex dipole. The results of numerical simulation presented in this section both confirm the leading-order analysis undertaken in previous sections, and provide crucial insight into the non-asymptotic regime by showing how the snail reliably emerges and stably evolves from typical initial conditions of the appropriate symmetry.

\subsection{Setup}
We simulate AFWOS subject to the following conditions:  vorticity is antisymmetric about the $z=0$ plane, and  vorticity is nonzero only in a small region (which possibly moves over time, and may lie far from the $r=0$ axis). The configuration of the simulation is represented at each time step by the values of $\theta$-vorticity within a small square region of the plane. Time stepping is implemented via the 4th order Runge--Kutta scheme, with the components of the Euler equation recovered from the vorticity via the Biot--Savart law. Calculations are undertaken in a local Fourier basis to preserve as much spatial accuracy as possible, both in the simulation, and in the computation of quantities of interest afterwards. While this Fourier basis has many convenient features  favouring the speed and accuracy of the simulation, there are a number of complications introduced by the mismatch between the periodic nature of the  basis and the infinite domain of the cylindrical coordinate system.  %We discuss how these are overcome in appendix B.

Every initial configuration we simulated evolved into a ``snail"; we here examine the trajectory of one configuration in depth. We describe the initial conditions here and explain why they could be expected to evolve into a snail in a particularly direct and smooth way. Specifically, the initial condition consists of two anti-parallel vortex rings, each of which has vorticity which is the product of the radial coordinate $r$, with a smooth transition function which is close to 1 inside a torus and close to 0 outside it: an appropriately shifted and scaled error function applied to the distance from circular center line of each torus. Specifically, given cylindrical coordinates $(z,r,\theta)$, the initial condition for the vorticity in the $\theta$ direction is defined by 
\[
\omega=r \,\operatorname{erf}\left(\frac{\sqrt{(r-0.7)^2+(z-0.32)^2}}{0.06}-3\right)-r\, \operatorname{erf}\left(\frac{\sqrt{(r-0.7)^2+(z+0.32)^2}}{0.06}-3\right) . 
\]

Thus our initial conditions have vortex rings at $(r,z)=(0.7,\pm 0.32)$, each of thickness roughly $3\times 0.06=0.18$, and with the transition from the interior to the exterior of each ring occurring over roughly 0.06 distance. The vorticity is chosen to be nearly homogeneous (before the $r$ scaling) inside each vortex ring so that when the snail sheds the outer layers of each ring, the vorticity will become increasingly constant inside the evolving snail. 

Starting with vortex rings at radius $0.7$, the simulation was run until the radius at the center of the snail was 9.6. The diameter of the vortex tube was initially $0.36$ and was finally 0.12
in the radial direction and 0.039 in the z direction, having shed 55\% of
its circulation.

Because our simulation repeatedly increases resolution so that the snail remains several hundred pixels across, the pixel size decreases from $0.002$ down to $0.00035$ over the course of a run, and would continue to decrease as $r$ increases. Stability concerns dictate that the evolving vortex rings can move at most a fraction of a pixel in each simulation time step, meaning that the cost of continuing for larger $r$ would continue to increase rapidly, and running over a much wider range of radii is infeasible. Nonetheless, we present results in Section~\ref{sec:numerical-results} showing that already within the scope of this simulation, the configuration converges rapidly to the expected behavior of the snail.

\subsection{Results}\label{sec:numerical-results}

\begin{figure} % float placement: (h)ere, page (t)op, page (b)ottom, other (p)age
  \centering
  % file name: C:/Users/Steve/Desktop/RESEARCH/Newton2012/newtonpaper/snail.eps
  \includegraphics[bb=-384 126 998 666,width=5.67in,height=2.21in,keepaspectratio]{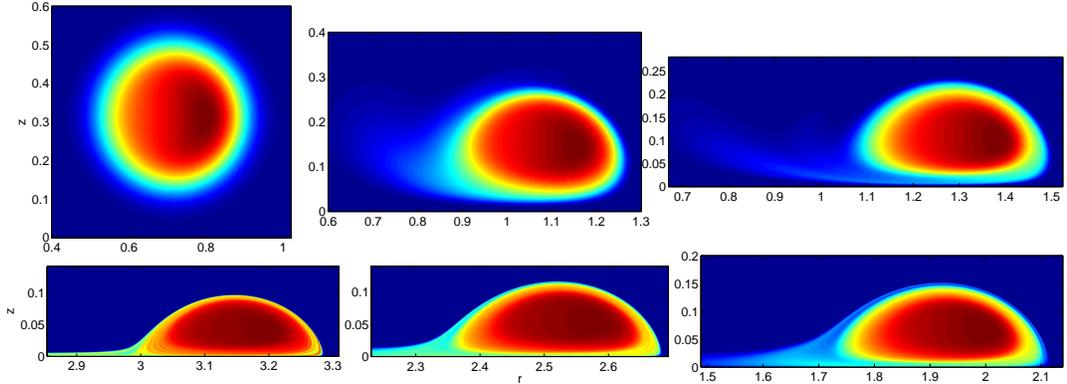}
  \caption{Development of the snail, shown at times t=0, 5.25, 9.5, 18, 25.}
  \label{fig:snail}
\end{figure}

We describe the results of the simulation here, both qualitatively and by quantitatively verifying scaling laws.
In Figure~\ref{fig:uneven} we show several snapshots: in each case, we are displaying a slice in the $(r,z)$ plane, depicting vorticity in the $\theta$ direction (note the changing scale of each panel). Initially, we have two vortex tubes, relatively diffuse, and separated from each other. They quickly move towards each other (without moving much away from the $z$-axis yet. When well separated the tubes are driven together by a converging flow along the binormal, as in the filament computations of \cite{PS}.  Soon, the tubes have essentially hit the $z=0$ symmetry plane and begin to shed vorticity into the tail as the dipole expands and the tubes are stretched. 
From here, the recognizable snail shape develops. 

The lateral extent of the vortex tubes decreases significantly as the tubes are stretched away from the axis and shed volume; the thickness of the shed tail also decreases relative to the thickness of the snail, since otherwise the snail would lose all its volume in finite time. 
While the speed of the snail increases with distance from the axis, the shed tail is essentially stationary, meaning that however thick the tail the snail sheds when at a certain
radius from the z axis, the shed tail essentially remains that thick at
that radius forever.

\begin{figure} % float placement: (h)ere, page (t)op, page (b)ottom, other (p)age
  \centering
  % file name: C:/Users/Steve/Desktop/RESEARCH/Newton2012/newtonpaper/radius-vs-time.eps
  \includegraphics[bb=-389 28 1002 763,width=5.67in,height=2.99in,keepaspectratio]{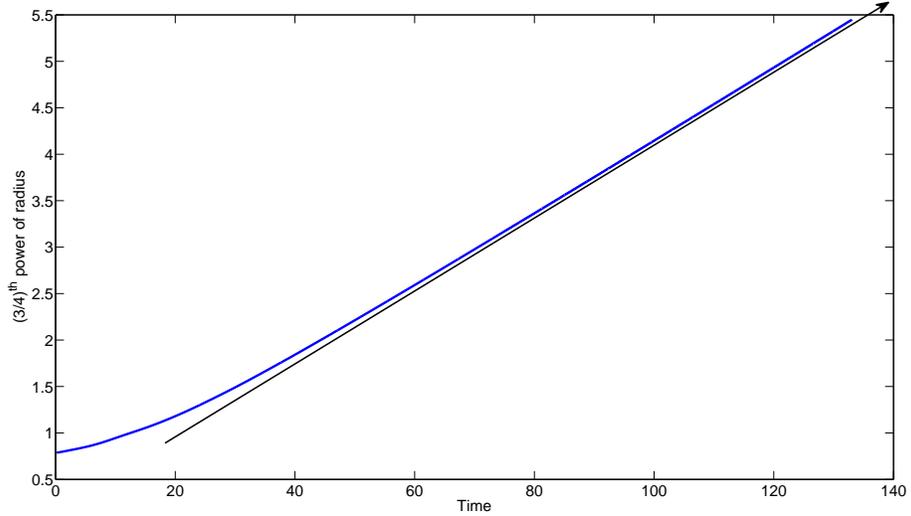}
  \caption{$R^{3/4}$ versus time, verifying a linear asymptote.}
  \label{fig:radius-vs-time}
\end{figure}

We next measure several aspects of the simulated snail and confirm that they follow the expected scalings. One of the key surprises of the snail is that its velocity increases without limit. Explicitly, the prediction is that the radius of each vortex ring should grow super-linearly with time, now adopting the dipole position $R(t)$ used earlier, as $$R\sim (t-c)^{4/3},$$ where the additive constant $c$ captures the fact that the start time of the simulation is arbitrary. To demonstrate this $4/3$-power relation, we instead plot in Figure~\ref{fig:radius-vs-time} the $3/4$-power of both sides, $R^{3/4}\sim t+c$: in blue is the radius of the center of each vortex tube as a function of time; drawn below in black is an arbitrary line to help  verify visually the claim that, asymptotically, $R^{3/4}\sim t+c$. Thus the snail does in fact accelerate over time, with radius proportional to $t^{4/3}$ and velocity proportional to its derivative, $t^{1/3}$.

\begin{figure} % float placement: (h)ere, page (t)op, page (b)ottom, other (p)age
  \centering
  % file name: C:/Users/Steve/Desktop/RESEARCH/Newton2012/newtonpaper/thickness-vs-radius.eps
  \includegraphics[bb=-389 19 1001 772,width=5.67in,height=3.07in,keepaspectratio]{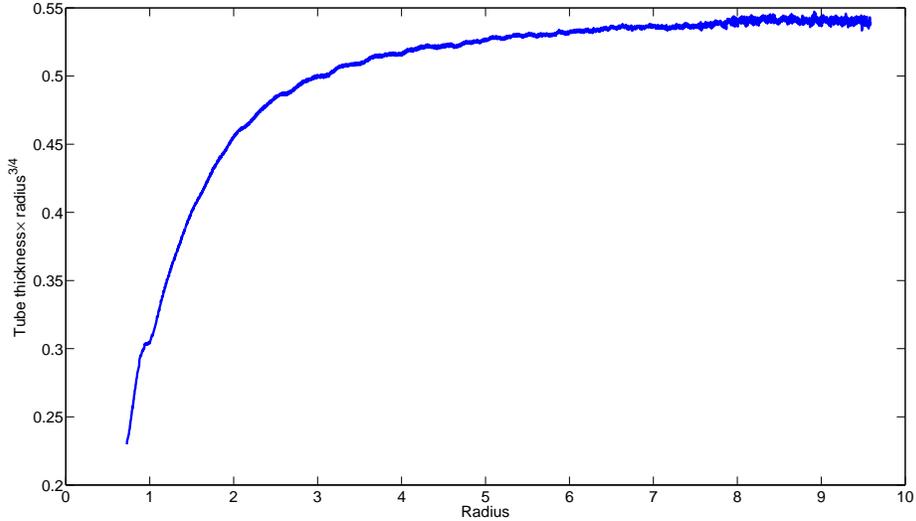}
  \caption{Dipole thickness times $R^{3/4}$ as a function of $R$.}
  \label{fig:thickness-vs-radius}
\end{figure}

The other main prediction of the snail model concerns how volume is shed, and how the  dimensions of the tube decrease faster than mere stretching would allow. Since we expect the thickness of the snail to decay as $R^{-3/4}$, we plot  in Figure~\ref{fig:thickness-vs-radius} 
the thickness in the $z$ direction times $R^{3/4}$; our result is that this quantity does indeed approach a constant as the simulation progresses.

\begin{figure} % float placement: (h)ere, page (t)op, page (b)ottom, other (p)age
  \centering
  % file name: C:/Users/Steve/Desktop/RESEARCH/Newton2012/newtonpaper/transition-vs-radius.eps
  \includegraphics[bb=-389 19 1001 772,width=5.67in,height=3.07in,keepaspectratio]{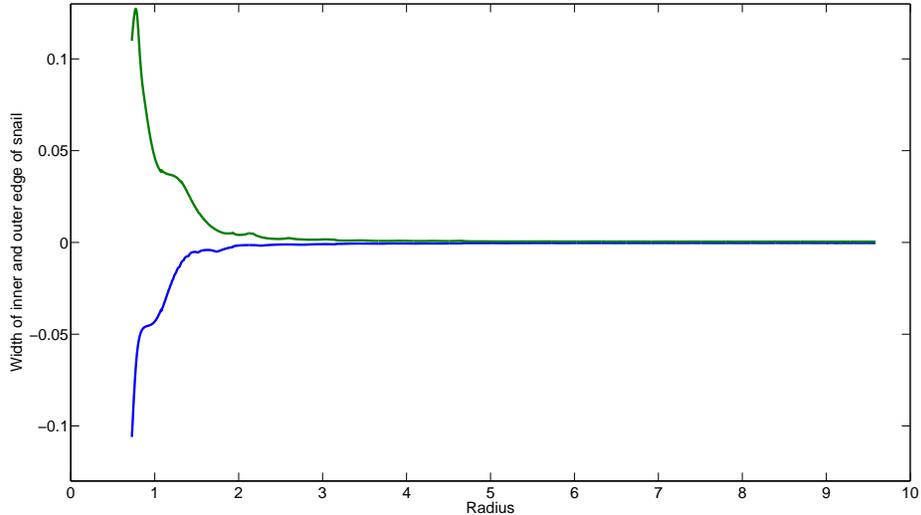}
  \caption{The width of the transition layer at the outer edge of the snail, versus $R$.}
  \label{fig:transition-vs-radius}
\end{figure}

Finally, one more qualitative prediction which is supported numerically is that the ``edge" of the snail, that is, the width of the transition from high vorticity to low vorticity, sharpens exponentially quickly with time.
We plot in Figure~\ref{fig:transition-vs-radius} the width of the fastest transition on the inner and outer edge of the snail (in the $z$-direction, computed using the inverse of the maximum derivative as a proxy for the width), and see that it rapidly falls to the order of a pixel size ($0.002$ initially, and decreasing over the course of the simulation).

\subsubsection{Other initial conditions}

It is revealing to consider the sensitivity of our dipole to small changes in the initial conditions. Two alternatives are especially worth discussing: first if the snail is inhomogeneous, how do ``lumps" in the snail translate to lumps in the tail, or affect the overall scaling? Secondly, how does the snail react to symmetry breaking, and in particular, violating the antisymmetry about the plane $z=0$?

\begin{figure} % float placement: (h)ere, page (t)op, page (b)ottom, other (p)age
  \centering
  % file name: C:/Users/Steve/Desktop/RESEARCH/Newton2012/newtonpaper/asymmetric.eps
  \includegraphics[bb=-33 1 645 791,width=5.67in,height=6.61in,keepaspectratio]{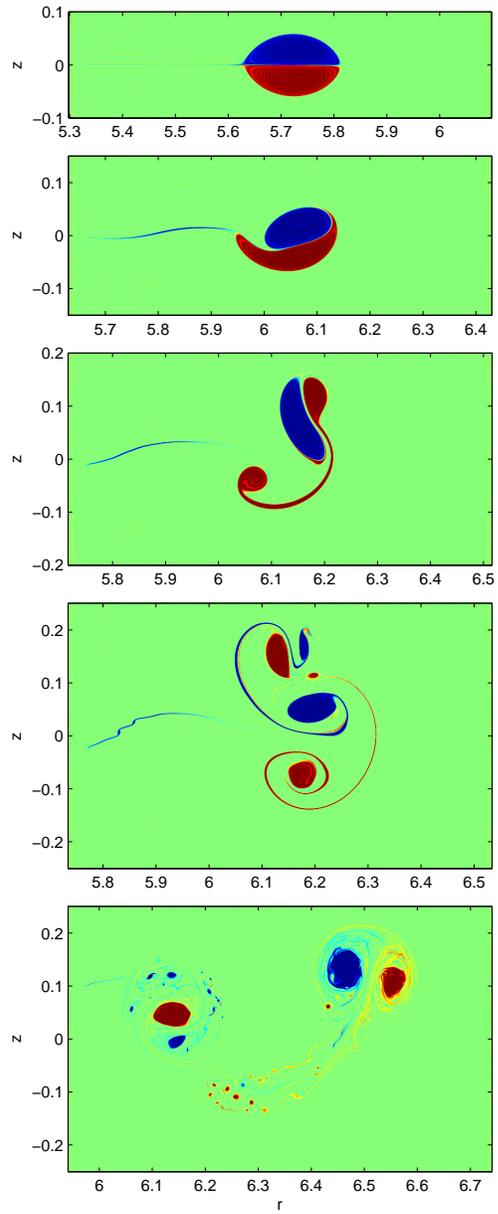}
  \caption{Breakup of the snail under asymmetric initial conditions. The vorticity in the upper eddy is ??? times that in the lower eddy.}
  \label{fig:asymmetric}
\end{figure}

The brief answers are that: variations of initial conditions that preserve symmetry do not much affect the snail, which appears to be a very robust phenomenon; however symmetry breaking  rapidly amplifies, leading to a breakdown of the dipole, where radial stretching not only stops accelerating, but typically stops entirely---thus the symmetries of the snail seem fundamental to its evolution. We show this breakup in Figure~\ref{fig:asymmetric}. It is significant that precise symmetry is needed to maintain vorticity growth, a point that is particularly important in the search for vortical structures which blow up in finite time. Also of interest in this example is the breakup of the dipole into two smaller dipole-like structures. 

\section {Discussion}

We have in this paper presented a model for vorticity growth in anti-parallel vortex structures in axisymmetric flow without swirl. The model provides an Euler flow which achieves the maximum possible growth of $|\omega|_{\max}$ as $t^{4/3}$.  The new feature of this work is the explicit role of vortex erosion, leading to scalings quite distinct from those associated with intact vortex tubes. The governing assumption behind the new scaling is the local conservation of energy following Lagrangian parcels of fluid. Our analysis has been restricted to a symmetric dipole arrangement of equal and opposite vortical eddies, which leads to the Sadovskii structure. Their might exist  asymmetric 2D dipoles of Sadovskii type, with constant vorticity in each eddy but differing circulations, which move on circular orbits.  However our calculations for the axisymmetric case indicate that the breaking of this symmetry leads to break-up of the dipole. This suggests that a hairpin singularity (the case $\beta = 4$ of \cite{Ch1}) would be highly unstable to the breaking of this symmetry, a will-o'-the-wisp in a turbulent flow. In essence extreme local amplification of vorticity seems to involve a delicate focusing of anti-parallel structures.

What are the implications of these calculations for more general Euler flows? The ``swirl'' which is absent in the present model amounts to flow in rings along the axis of the dipole. Axisymmetric flow {\em with} swirl can, according to \cite{HL}, blow up in finite time in the presence of an impenetrable boundary. In 
$ \scriptstyle{\mathbb{R}}^3$ the situation is unclear. In general three-dimensional dipole models must cope with the generation of axial flow by the axial pressure gradient  produced as the dipole stretches differentially. Part II will apply many of the ideas of the present paper to the ``hairpin'' geometry and discuss the possible role of axial flow on the resulting growth of vorticity.

\appendix

We thank Eric Siggia for his interest in this work and for his helpful comments.

\section{Potential flow past an expanding torus { of constant volume}}

A  torus of radius $R$ and cross-sectional area $\pi a^2$ expands radially (i.e. outward in the plane of symmetry) in a perfect inviscid fluid, $R=R(t)$, while maintaining a constant volume. What is the resulting irrotational flow field?

We first consider the potential
\bg
\phi= -{R\over 4\pi}\int_0^{2\pi}{d\theta\over \sqrt{R^2+r^2-2Rr\cos 2\theta +z^2}} \, ,
\ela{basepot}
\ee
representing a uniform distribution of sources  over the circle $z=0$, $r=R$. This can be brought into the form
\bg
\phi = -{R\over \pi}\int_0^{\pi/2} {d\theta\over\sqrt{(R+r)^2+z^2-4Rr\sin^2\theta}}\, ,
\ee
or
\bg
\phi= -{R\over \pi P} \int_0^{\pi/2}{d\theta\over\sqrt{1-k^2\sin^2\theta}}= -{R\over \pi P}\;K(k),
\ee
where
\bg 
P=\sqrt{(R+r)^2+z^2},\quad k^2={4Rr\over  (R+r)^2+z^2}\, .
\ee
Near $k=1$ we have ( \cite{CG}) 
\bg
K(k)=\sum_{n=0}^{N-1}\bigg[{({1\over 2})_n\over n!}\bigg]^2\left[ \log{1\over k^\prime}+\psi(1+n)-\psi(1/2+n)\right] (k^\prime)^{2n}+O(k^\prime)^{2N}\log k^\prime,
\ee
where
\bg
\big(\tfrac{1}{2}\big)_n={\Gamma(\big(n+{1\over 2}\big)\over \Gamma\big({1\over 2}\big)} \, ,
\quad
%\ee
%\bg
\psi(1-\psi(1/2)=2\log 2,
\ee
\bg
\psi(1+n)-\psi(1/2+n)=2\Big[ \log 2-1+{1\over 2}-\dots-{1\over 2n-1}+{1\over 2n}\Big],\quad
n\geq 1,
\ee
and
\bg
k^\prime= \sqrt{1-k^2}\, .
\ee

Going over to local coordinates we have $r=R+x$, $z = y$, $\rho^2=x^2+y^2$. Then we have

\bg
P=2R\sqrt{1+x/R+ \tfrac{1}{4} \rho^2/R^2}\, ,\quad 
k^\prime = {\rho / P}  \, . 
\ee

Expanding through terms of order $(\rho/R)^2$ we have
\bg
\phi= -{1\over 2\pi}\left[1-{1\over 2} {x\over R}+{3\over 8}\left({x\over R}\right)^2-{1\over 8}\left({\rho\over R}\right)^2
\right] \left[\log{8R\over \rho}+{1\over 2} {x\over R} +{9\over 4} \bigg(\log{8R\over \rho}-1\bigg){\rho^2\over 4R^2}\right]+ o\left({\rho^2\over R^2}\right).
\ee
This gives the ordering
\bg
2\pi\phi= -\log{8R\over \rho}+\bigg[{x\over 2}\Big(\log{8R\over \rho}-1\Big)\bigg]{1\over R} + \cdots.
\ela{lead}
\ee
We will use these terms in the expansion of $\phi$ to solve the problem of the torus of constant volume.

We seek the potential flow past a torus expanding so that $R(t)$ increases with time, with the radius $a(t)$ of the cross section satisfying
\bg
{\dot{a}\over a } = -{1\over 2} \, {\dot{R}\over R} \, .
\ela{shrink}
\ee

We use the fact that if $\phi$  solves Laplace's equation in 3D, then so does 
$R\phi^\prime = x\phi_x+y\phi_y+z\phi_z$ or, with radial symmetry,
\bg
R\phi^\prime = r\phi_r+ z\phi_z= R\phi_x+\rho \phi_\rho,
\ee
or
\bg 
\phi^\prime=\phi_x+R^{-1}\rho\phi_\rho.
\ela{new}
\ee
Using \er{lead} for the expansion of $\phi$ in \er{new} we see that
\bg
2\pi \phi^\prime\equiv \Phi +{1\over R} ={x\over \rho^2}+{1\over 2R}\log{8R\over\rho}-{1\over 2R} {x^2\over \rho^2}+{1\over R}+O(R^{-2}).
\ee
Thus $\Phi$ is a building block of the local potential for a cylindrical cross-section. Indeed
\bg
-\dot{R} (x+a^2\Phi)\sim -\dot{R}x (1+a^2/\rho^2)
\ee
is the potential for uniform flow over a cylinder.

Now in the neighbourhood of infinity we see that 
\bg
\phi^\prime\sim {1\over 2\sqrt{r^2+z^2}} \, ,
\ee
giving a net source flux of $-2\pi$. We can check that this is consistent with flux out of the surface of the torus. Indeed
$ x/ (2\pi \rho^2)$ contributes
\bg
 {1\over 2\pi}\int_0^{2\pi} (-\cos\theta/\rho^2)2\pi{\rho}(R+\rho\cos\theta)\, d\theta=-\pi,
\ee
and $(4\pi{R})^{-1} \log ( 8R / \rho) $ contributes
\bg
{2\pi R\over 4\pi R}\int_0^{2\pi} (-1/\rho)\rho \, d\theta = -\pi.
\ee
To obtain a potential free of net source strength we must then add on $\phi/R$, and so  the potential of the expanding torus, at the point where its cross-sectional radius is $a$, relative to the fluid at infinity (not comoving), is
\bg
\phi_{\mathrm{torus}} = -2\pi \dot{R}  a^2 (\Phi + \phi/R) \sim a^2\dot{R}\left[ -{x\over \rho^2}+{1\over 2 R} \log {8R\over \rho} +{1\over 2R} {x^2\over \rho^2}\right]+O(a^2\dot{R}/R^2).
\ela{finphi}
\ee
Recalling \er{shrink}, it is readily seen that the exhibited terms lead to the appropriate normal velocity at the instantaneous surface of the torus.

%\section{Implementation details for the simulations of section 5.}

\bibliographystyle{jfm}

\end{document}